
\input harvmac

\def\IR{\relax{\rm I\kern-.18em R}}
\def\no{\noindent}

\def\qq{\qquad}
\def \L {{\cal L}}
\def \bab {{\bar b }}
\def \tu {{\tilde u}}
\def \diag  {{\rm diag }}
\def \dim  {{\rm dim \ }}
\def \bd {\bar \del}

\def \tt {{\tilde \t}}
\def \G {\Gamma}\def \tr { {\tilde \r}} \def \k1 {{1\over
k}} \def \bh { {\bar h} } \def \ov { \over }

\def \O {\Omega }

\def \ra {\rightarrow}
\def \HL {{H_L}}
\def \HR {{H_R}}

\def \W { {\bar \o }}
\def \half{{\textstyle{1\over 2}}}

\def \a {\alpha}
\def \b {\beta}

\def \Tr {{\ \rm Tr \ }}

\def \ln {{\rm \ ln \  }}
\def \det {{\ \rm det \ }}

\def \1p {{1\over  \pi }}
\def \2p {{{1\over  2\pi }}}
\def \4p {{ {1\over 4 \pi }}}
\def \8p {{{1\over 8 \pi }}}
\def \P^* { P^{\dag } }
\def \p {\phi}

\def \ep {\epsilon}

\def \r {\rho}
\def \k {\kappa }
\def \d {\delta}
\def \o {\omega}
\def \s {\sigma}
\def \t {\theta}

\def \third {{\textstyle{1\over 3}}}

\def \e#1 {{{\rm e}^{#1}}}

\def \eq#1 {\eqno {(#1)}}
\def \sm {$\s$-model\ }

\def \bd  {{ \bar \del }}

\def \bd  { \bar \del }

\def \ov {\over }
\def \kg { k + {\rm g_G}}
\def \kh { k + {\rm g_H}}
\def \A  { {\bar A} }
\def \tth {\tilde h}

\def \H {{\cal H}}
\def \o {\omega}

\def \p {\phi}
\def \ep {\epsilon}
\def \s {\sigma}

\def \gr {\rho}
\def \r {\rho}
\def \d {\delta}

\def \third {{1\over 3}}
\def \e#1 {{{\rm e}^{#1}}}

\def \hg {{\hat g}}

\def \H {{\cal H}}

\def \tg {{\tilde g}}
\def \half { { 1\over 2 }}

\def \J {\bar J }
\def\np {  Nucl. Phys. }
\def \pl { Phys. Lett. }
\def \mpl { Mod. Phys. Lett. }
\def \prl { Phys. Rev. Lett. }
\def \pr  { Phys. Rev. }
\def \ap  { Ann. Phys. }
\def \cmp { Commun. Math. Phys. }
\def \ijmp { Int. J. Mod. Phys. }
\baselineskip8pt
\Title{\vbox
{\baselineskip6pt{\hbox{CERN-TH.6962/93}}\hbox{USC-93/HEP-S2}{\hbox
{Imperial/TP/92-93/65}}{\hbox{hep-th/9308018}} }}  {\vbox{\centerline { Chiral
gauged WZNW models
}\vskip2pt
 \centerline{   and heterotic string  backgrounds    }
}}
\vskip -34 true pt
\centerline  { {  Konstadinos Sfetsos }}
\vskip 1pt
\centerline {\it Physics Department, University of Southern California }
\centerline {\it Los Angeles, CA 90089-0484, USA}

\centerline  {\it and}
\vskip 1pt
\centerline {\it  Institute for Theoretical Physics, Utrecht University}
\centerline {\it  Princetonplein 5, TA 3508, The Netherlands }
\vskip 3pt
\centerline {and}
\vskip 3pt
\centerline{   A.A. Tseytlin\footnote{$^{*}$}{\baselineskip5pt
On leave  from Lebedev Physics
Institute, Moscow, Russia.
e-mail: tseytlin@surya3.cern.ch and tseytlin@ic.ac.uk} }
\vskip 1pt
\centerline {\it Theory Division, CERN}
\centerline {\it
CH-1211 Geneva 23, Switzerland}

\centerline  {\it and}

\centerline {\it  Blackett Laboratory, Imperial College}
\centerline {\it London SW7 2BZ, U.K.}
\vskip 3pt
\centerline {\bf Abstract}
\vskip 3pt
\baselineskip6pt
\noindent
We construct new heterotic string backgrounds  which are analogous
to superstring solutions corresponding to coset models but are not
simply the `embeddings'of the latter. They are described by the
(1,0) supersymmetric extension of the $G/H$ chiral  gauged WZNW
models. The `chiral gauged' WZNW action differs from the standard
gauged WZNW action by the absence of the  $A\bar A$-term (and thus
is not gauge invariant in the usual sense) but can still be expressed
as a combination of WZNW actions and is conformal invariant. We explain
a close relation between gauged and chiral gauged  WZNW models
and  prove that in the  case of the abelian $H$ the $G/H$ chiral gauged
theory is equivalent to a particular $(G\times H)/H$ gauged WZNW theory.
In contrast to the gauged  WZNW model, the chiral gauged one admits a (1,0)
supersymmetric extension which is consistent at the quantum level.
Integrating out the $2d$ gauge field we determine the exact (in $\alpha'$)
form  of the couplings  of the corresponding heterotic sigma model. While
in the bosonic (superstring) cases all the fields  depend (do not depend)
non-trivially on  $\alpha'$ here the metric receives only one $O(\alpha')$
correction  while the antisymmetric tensor and the dilaton remain
semiclassical.
As a simplest example, we discuss the basic $D=3$ solution which is the
heterotic
string counterpart of the `black string' $SL(2,R) \times R/ R $ background.
\vskip 2pt
\noindent
{CERN-TH.6962/93}

\noindent
{August 1993}
\Date { }

\noblackbox
\overfullrule=0pt
\baselineskip 20pt plus 2pt minus 2pt

\lref \bep {C. Becchi and O. Piguet, \np B315(1989)153. }
\lref \nov {S.P. Novikov,  Sov. Math. Dokl. 37(1982)3. }
\lref \fuch { J. Fuchs, \np B286(1987)455 and  B318(1989)631. }
\lref \sus { R. Rohm,  \pr D32(1985)2845.}
\lref \suss {   H.W. Braden, \pr D33(1986)2411.}
\lref   \red { A.N. Redlich and H.J. Schnitzer, \pl B167(1986)315 and
B193(1987)536(E);  A. Ceresole,
 A. Lerda, P. Pizzochecco
 and
P. van Nieuwenhuizen, \pl
 B189(1987)34.}
 \lref \div  { P. Di Vecchia,  V. Knizhnik, J. Peterson and P. Rossi, \np
B253(1985)701.}

\lref \schn {  H. Schnitzer, \np B324(1989)412.  }

\lref \all { R.W. Allen, I. Jack and D.R.T. Jones, Z. Phys. C41(1988)323. }

\lref \nem {D. Nemeschansky and S. Yankielowicz, \prl 54(1985)620; 54(1985)1736
(E).}

\lref \bep { C. Becchi and O. Piguet, \np B315(1989)153. }

\lref \ks {Y. Kazama and H. Suzuki, \np B321(1989)232; \pl B216(1989)112.}

\lref \mor {A.Yu. Morozov, A.M. Perelomov, A.A. Rosly, M.A. Shifman and A.V.
Turbiner, \ijmp
A5(1990)803.}

 \lref \tur {A.V. Turbiner, \cmp 118(1988)467;  M.A. Shifman and A.V. Turbiner,
\cmp 126(1989)347;
M.A. Shifman, \ijmp A4(1989)2897.}

\lref \hal { M.B. Halpern and E.B. Kiritsis, \mpl A4(1989)1373  and
A4(1989)1797
(E).}

\lref \haly {M.B. Halpern and   J.P. Yamron,  Nucl.Phys.B332(1990)411;
Nucl.Phys.
B351(1991)333.}
\lref \halp { M.B. Halpern, E.B. Kiritsis, N.A. Obers, M. Porrati and J.P.
Yamron,
\ijmp A5(1990)2275;
  A.Yu. Morozov,  M.A. Shifman and A.V. Turbiner, \ijmp
A5(1990)2953;
A. Giveon, M.B. Halpern, E.B. Kiritsis and  N.A. Obers,
\np B357(1991)655.}
\lref \bpz {A.A. Belavin, A.M. Polyakov and A.B. Zamolodchikov, \np
B241(1984)333. }
\lref \efr {     S. Elitzur, A. Forge and E. Rabinovici, \np B359 (1991)
581;
 G. Mandal, A. Sengupta and S. Wadia, Mod. Phys. Lett. A6(1991)1685. }
\lref \sak {K. Sakai, Kyoto preprint, KUNS-1141-1992. }
\lref \ver {H. Verlinde, \np B337(1990)652.}
\lref \gwz {      K. Bardakci, E. Rabinovici and
B. S\"aring, \np B299(1988)157;
 K. Gawedzki and A. Kupiainen, \pl B215(1988)119;
\np B320(1989)625. }

\lref \sen {A. Sen, preprint TIFR-TH-92-57. }

\lref \bcr {K. Bardakci, M. Crescimanno and E. Rabinovici, \np
B344(1990)344. }
\lref \Jack {I. Jack, D.R.T.  Jones and J. Panvel,  \np B393(1993)95. }
\lref \zam  { Al. B. Zamolodchikov, preprint ITEP 87-89. }

\lref \hor {J. Horne and G. Horowitz, \np B368(1992)444. }
\lref \tse { A.A. Tseytlin, \pl B264(1991)311. }
\lref \GWZNW  { P. Di Vecchia and P. Rossi, \pl  B140(1984)344;
 P. Di Vecchia, B. Durhuus  and J. Petersen, \pl  B144(1984)245.}
\lref \oal { O. Alvarez, \np B238(1984)61. }

\lref \ishi { N. Ishibashi, M.  Li and A. Steif, \prl 67(1991)3336. }
\lref  \kuma  { M. Ro\v cek and E. Verlinde, \np B373(1992)630; A. Kumar,
preprint CERN-TH.6530/92;
 S. Hussan and A. Sen,  preprint  TIFR-TH-92-61;  D. Gershon,
preprint TAUP-2005-92; X. de la Ossa and F. Quevedo, preprint NEIP92-004; E.
Kiritsis, preprint LPTENS-92-29. }

\lref \rocver { A. Giveon and M. Ro\v cek, \np B380(1992)128. }
\lref \frts {E.S. Fradkin and A.A. Tseytlin, \np B261(1985)1. }
\lref \mplt {A.A. Tseytlin, \mpl A6(1991)1721. }
\lref\bn {I. Bars and D. Nemeschansky, \np B348(1991)89.}
\lref \shif { M.A. Shifman, \np B352(1991)87.}
\lref\wittt { E. Witten, \cmp 121(1989)351; G. Moore and N. Seiberg, \pl
B220(1989)422.} \lref \chernsim { E. Guadagnini, M. Martellini and M.
Mintchev, \np B330(1990)575;
L. Alvarez-Gaume, J. Labastida and A. Ramallo, \np B354(1990)103;
G. Giavarini, C.P. Martin and F. Ruiz Ruiz, \np B381(1992)222; preprint
LPTHE-92-42.}
\lref \shifley { H. Leutwyler and M.A. Shifman, \ijmp A7(1992)795. }
\lref \polwig { A.M. Polyakov and P.B. Wiegman, \pl B131(1984)121; \pl
B141(1984)223.  }
\lref \polles { A. Polyakov, in: {\it Fields, Strings and Critical Phenomena},
  Proc. of Les Houches 1988,  eds.  E. Brezin and J. Zinn-Justin
(North-Holland,1990).   }
\lref \kutas {
D. Kutasov, \pl B233(1989)369.} \lref \karabali { D. Karabali, Q-Han Park, H.J.
Schnitzer and
Z. Yang, \pl B216(1989)307;  D. Karabali and H.J. Schnitzer, \np B329(1990)649.
}
\lref \ginq {P. Ginsparg and F. Quevedo,  \np B385(1992)527. }
\lref \gko  {K. Bardakci and M.B. Halpern, \pr D3(1971)2493;
 M.B. Halpern, \pr D4(1971)2398;   P. Goddard,
A. Kent and D. Olive, \pl B152(1985)88; \cmp 103(1986)303.}
%
\lref \dvv  { R. Dijkgraaf, H. Verlinde and E. Verlinde, \np B371(1992)269. }
\lref \kniz {  V. Knizhnik and A. Zamolodchikov, \np B247(1984)83. }

\lref \witt { E. Witten, \cmp 92(1984)455.}
\lref \wit { E. Witten, \pr D44(1991)314.}
\lref \anton { I. Antoniadis, C. Bachas, J. Ellis and D.V. Nanopoulos, \pl B211
(1988)393.}
\lref \bsfet {I. Bars and  K. Sfetsos, \pr D46(1992)4510.}
%
%
\lref \ts  {A.A.Tseytlin, \pl B268(1991)175. }
\lref \kir {{E. Kiritsis, \mpl A6(1991)2871.} }

\lref \shts {A.S. Schwarz and A.A. Tseytlin,  preprint Imperial/TP/92-93/01
(1992). }
\lref \bush { T.H. Buscher, \pl B201(1988)466.  }

\lref \plwave { D. Amati and C. Klim\v cik, \pl B219(1989)443; G. Horowitz and
A. Steif, \prl 64(1990)260.}

\lref\bsft { I. Bars, preprint USC-91-HEP-B3. }

\lref\bs { I. Bars and K. Sfetsos,  preprint USC-93/HEP-B1 (1993);
Phys. Rev. D (1993) }
\lref\bst { I. Bars,  K. Sfetsos and A.A. Tseytlin, unpublished. }
\lref\bb  { I. Bars, \np B334(1990)125. }
\lref \tsw  { A.A. Tseytlin,\np B399(1993)601.}
\lref \ger { A. Gerasimov, A. Morozov, M. Olshanetsky, A. Marshakov and S.
Shatashvili, \ijmp
A5(1990)2495. }
\lref \hull {C.M. Hull and B. Spence, \np B345(1990)493. }

\lref \sch { K. Schoutens, A. Sevrin and P. van Nieuwenhuizen,
in: Proc. of the Stony Brook Conference {\it `Strings and Symmetries 1991'},
p.558  (World
Scientific, Singapore, 1992).} \lref \boer { J. de Boer and J. Goeree, Utrecht
preprint THU-92/33. }
\lref \dev { C. Destri and H.J. De Vega, \pl B208(1988)255. }
\lref  \noj {S. Nojiri, \pl B271(1992)41. }
\lref \swz  { E. Witten, \np B371(1992)191;
T. Nakatsu, Progr. Theor. Phys. 87(1992)795. }
 \lref \bsf { I. Bars and K. Sfetsos, \pl B277(1992)269. }

\lref \br {A. Barut and R. Raczka, ``Theory of Group Representations and
Applications", p.120
 (PWN, Warszawa 1980). }
\lref \jjmo {I. Jack, D.R.T. Jones, N.Mohammedi and H. Osborn, \np
B332(1990)359;
C.M. Hull and B. Spence, \pl B232(1989)204. }

\lref \tttt { A.A. Tseytlin,  preprint Imperial/TP/92-93/7 (1992), \pr
D47(1993) no.8.}
\lref \per { V. Novikov, M. Shifman, A. Vainshtein and V. Zakharov, \pl
B139(1984)389;
A. Morozov, A. Perelomov and M. Shifman, \np B248(1984)279;
M.C. Prati and A.M. Perelomov, \np B258(1985)647. }
\lref \alv {
L. Alvarez-Gaum\'e,   D. Freedman and S. Mukhi, \ap 134(1981)85;
L. Alvarez-Gaum\'e, \np B184(1981)180. }
\lref \gr {  M.T. Grisaru, A. van de Ven and D. Zanon, \np B277(1986)409. }
\lref \call {C.G. Callan, D. Friedan, E. Martinec and M.J. Perry,
Nucl. Phys.B262(1985)593; A. Sen, Phys. Rev. D32(1985)316; \prl 55(1985)1846.}
\lref \hul {C.M. Hull, \pl B167(1986)51; Nucl. Phys. B267(1986)266.}
\lref \het { Y. Cai and C.A. Nunez, \np B287(1987)279;
  Y. Kikuchi and  C. Marzban, \pr D35(1987)1400.}
\lref \met {R.R.Metsaev and A.A.Tseytlin,\pl B185(1987)52;\np B293(1987)385.}
\lref \gro {D.J. Gross and J.H. Sloan, \np B291(1987)41. }
\lref  \foa { A.P. Foakes, N. Mohammedi and D.A. Ross, \pl B206(1988)57;
\np B310(1988)335.}
\lref \hw {C.M. Hull and E. Witten, \pl B160(1985)398.}
\lref \hh { D.J. Gross, J.A. Harvey, E. Martinec and R. Rohm, \np B256(1985)253
and  B267(1986)75.}

\lref \chsw { P. Candelas, G. Horowitz, G. Strominger and E. Witten, \np
B258(1985)46. }
\lref \aat { A.A. Tseytlin, preprint CERN-TH.6804/93.}
\lref \at {A.A. Tseytlin, preprint CERN-TH.6820/93. }
\lref \aps { S. De Alwis, J. Polchinski and R. Schimmrigk, \pl B218(1989)449. }
\lref \gny  { M.D. McGuigan, C.R. Nappi and S.A. Yost, \np B375(1992)421. }
\lref \py  {  M.J.  Perry and E. Teo, preprint DAMTP R93/1 (1993); P. Yi,
preprint CALT-68-1852
(1993). }
\lref \tv {   A.A. Tseytlin and C. Vafa, \np B372(1992)443.}
\lref \hull {C.M. Hull and B. Spence, \np B345(1990)493. }
\lref \gep { D. Gepner, \pl B199(1987)370; \np B296(1988)757.}
\lref \gat {S.J. Gates, S.V. Ketov, S.M. Kuzenko and O.A. Soloviev, \np
B362(1991)199.}
\lref \ell {U. Ellwanger, J. Fuchs and M.G. Schmidt, \pl B203(1988)244; \np
B314(1989)175. }
\lref \ket { S.V. Ketov and O.A. Soloviev, \pl B232(1989)75; \ijmp
A6(1991)2971. }
\lref \GR {A. Giveon and M. Ro\v{c}ek, Nucl. Phys. B380(1992)128.}
\lref \giv {  A. Giveon, Mod. Phys. Lett. A6(1991)2843.}
\lref \GRV {A. Giveon, E. Rabinovici and G. Veneziano, Nucl. Phys.
B322(1989)167;
A. Shapere and F. Wilczek, \np B320(1989)669.}
\lref \GMR {A. Giveon, N. Malkin and E. Rabinovici, Phys. Lett. B238(1990)57.}
\lref \GS  {  A. Giveon and D.-J. Smit. Nucl. Phys.  B349(1991)168.}
\lref  \GP   {  A. Giveon and A. Pasquinucci, Phys. Lett. B294(1992)162.}
\lref \GG  { I.C. Halliday, E. Rabinovici, A. Schwimmer and M. Chanowitz, \np
B268(1986)413. }
\lref \Ve    { K.M. Meissner and G. Veneziano, \pl B267(1991)33;
M. Gasperini, J. Maharana and G. Veneziano, \pl B272(1991)277;
A. Sen,  \pl  B271(1991)295.}
\lref \eng { F. Englert, H. Nicolai and A.N. Schellekens, \np B274(1986)315.}
\lref \ler  {W. Lerche, D. L\"ust and A.N. Schellekens, \np B287(1987)477; \pl
B187(1987)45.}
\lref \fre  { S. Ferrara and P. Fre', \ijmp 5A(1990)989;
M. Billo', P. Fre', L. Girardello and A. Zaffaroni, preprint SISSA/159/92/EP.}
\lref \tye { S.-W. Chung and S.-H. Tye, \pr D47(1993)4546.}
\lref  \GRT   { A. Giveon, E. Rabinovici and A.A. Tseytlin,  preprint
CERN-TH.6872/93,
RI-150-93, hep-th/9304155.}
 \lref \kz {  V.G. Knizhnik and A.B. Zamolodchikov, \np B247(1984)83. }
\lref \sfet { K. Sfetsos, preprint USC-93/HEP-S1, hep-th/9305074.}
\lref  \horne { J.H. Horne and G.T. Horowitz, \np B368(1992)444.}

\lref \nap { M. Henningson and C. Nappi,  preprint IASSNS-HEP-92/88.  }

\lref \givkir {E. Kiritsis, preprint CERN-TH.6816/93; A. Giveon and E.
Kiritsis,  preprint
CERN-TH.6816/93, RIP-149-93. }
\lref \kumar { S.K. Kar and A. Kumar, \pl B291(1992)246.}
\lref \maha { S. Mahapatra, \mpl A7(1992)2999;
S.K. Kar, S.P. Khastrig and G. Sengupta, \pr D47(1993)3643;
S.K.  Kar, A. Kumar and G. Sengupta, preprint IP/BBSR/92-67.}
\lref \bow  {P. Bowcock, \np B316(1989)80. }
\lref \witten { E. Witten,  Commun. Math. Phys. 144(1992)189. }
\lref \sfts {K. Sfetsos and A.A. Tseytlin, preprint CERN-TH.6969, to appear.}
\lref \gin  { P. Ginsparg and F. Quevedo, \np B385(1992)527.}
\lref \sft    {K. Sfetsos, \np B389(1993)424.}
\lref\IBCS{ I. Bars,
in: {\it  Proc.
  of the  XX$^{th}$ Int. Conf. on Diff. Geometrical Methods in Physics}, eds.
S.
 Catto and A. Rocha, Vol. 2, p. 695 (World Scientific, 1992).}

\newsec {Introduction }

The Wess-Zumino-Novikov-Witten (WZNW) model \witt\nov\  is a prototypical
example of a local  field
theory which is conformally invariant, i.e.
which   provides a realisation of the  conformal operator algebra \kz.
The $G/H$ gauged  WZNW model (or GWZNW)  can be  represented   as a particular
combination of the WZNW models for a group $G$ and a subgroup $H$.
This explains its
conformal invariance as well as its  relation \gwz\karabali\ to a  coset
conformal field theory
\gko. Starting  with GWZNW  it is possible to construct various string
solutions
to the leading order in $\a'$
(see, for example, \wit
\ref\BSthree
{I. Bars and K. Sfetsos, \mpl  A7(1992)1091;
\pr D46(1992)4495.}\gin
\ref\NAWIT{C. Nappi and E. Witten, \pl B293(1992)309.}), or to all orders in
$\a'$
(see, for example, \dvv\bsfet\sft).

One may question  if there are other  non-trivial combinations of WZNW models
which   also correspond to local conformal field theories  and thus generate
new string  backgrounds.
A closely related to GWZNW model is the so called chiral gauged WZNW
model \tye\ (CWZNW) in which one does not
include the `counterterm' $\Tr(A\A)$  in the action \gin\
($A_m= (A, \A)$ is a $2d$ vector  field with values in the algebra of $H$)
and thus one has  only `chiral' gauge invariance (with gauge parameters
constrained to be holomorphic or antiholomorphic).
As we shall see below, the action of CWZNW  is, in
fact, the only
  local modification  of the GWZNW  action which  can also   be represented  in
terms of a
combination of independent WZNW
actions  and thus is  certainly  conformally invariant at the quantum level.
As a consequence,
the $\s$-model for the  coordinates of the group  space ($G$)  obtained from
a CWZNW model by integrating out  the $2d$ gauge field
(see \kumar\maha\ for particular examples  and \sfet\ for a general case)
will also be conformally invariant.

An interpretation that can be  given to a  `product' of WZNW models
depends crucially on which  combinations of fields are treated as `fundamental'
and which -- as
`auxiliary' or Lagrange multiplier-type  variables.  Since the actions of both
GWZNW and CWZNW  are
local,
 being  expressed  in terms of the  group variable $g$ and the $2d$  vector
field $A_m$
(with the latter  having  no kinetic term), it is natural to treat $A_m$ as an
auxiliary field
which should be integrated out (without introducing a source term
for it) in the path integral.
It is within such an  approach that the GWZNW model
is related to the coset model.
Below we shall study  a  possibility to give a similar
interpretation to CWZNW.\foot {We shall mostly
consider  CWZNW models in the  cases when the `left' and `right' subgroups
are the same.
We disagree with the claim \tye\ that the CWZNW model with $H_L=H_R=H$ is
equivalent to $G/H$ or GWZNW model. }
We shall find that  if  all the fields  are treated on an equal footing,
i.e. if  one admits a
possibility of making field redefinitions that mix $g$ with $A_m$,  then
the CWNW models are essentially equivalent to a subclass of GWZNW models:
  $(G/H)_{CWZNW}$   can be  identified (modulo field redefinitions)
with the   $(G/H \times H)_{GWZNW}$.
 When $H$ is abelian it is possible to  establish a more direct relation
without  the necessity to redefine $A_m$:
$(G/H)_{CWZNW}$  = $[(G\times H)/H]_{GWZNW}$, where in the GWZNW case
the subgroup $H$ is embedded into  $G\times H$  in a specific way
and  is gauged axially.
The latter  equivalence  provides a general explanation for  the
observations in \kumar\maha\ (to the leading order in $1/k$) and \sfet\
(exactly in $1/k$)  that
the $SL(2,\IR)/\IR$ CWZNW
model  is   a particular limit   of  the $SL(2,\IR)\times \IR /\IR $ GWZNW (or
`black string'
\horne\sft) model.

Our interest in the CWZNW models,
besides their importance as bosonic models with exact conformal invariance
(and in   connection  with  $O(d,d)$ duality
which seems to relate different `mixtures' of WZNW models,
see e.g. \GR\nap\givkir)
was originally motivated by   a desire
to understand if they can be used for a construction of   heterotic
string  backgrounds   which  are not simply the embeddings  of $(1,1)$
 supersymmetric coset solutions.  As  we
shall show,  this is indeed the case:  CWZNW model  has a  consistent  $(1,0)$
supersymmetric generalisation and    may thus   serve as a basis  for a
non-trivial   heterotic string world sheet theory.

The  question about heterotic string solutions related to coset models was
recently discussed in
\GRT\ where it was pointed out that the direct (1,0) truncation of the (1,1)
supersymmetric
gauged WZNW model  does not correspond to a consistent heterotic string
background  since the
resulting $2d$ theory is anomalous
(the  fermions couple chirally to the $2d$ gauge field which is
integrated over in the path integral).
It was suggested  to  cancel the $2d$ gauge
anomaly by introducing  an  additional world sheet  coupling term
which corresponds to a non-vanishing target space  gauge field  background
\GRT.
As a result, the theory  becomes effectively (1,1)
supersymmetric  and can be interpreted as a  coset  model superstring
solution embedded into the
heterotic string theory.
Because of (1,1) supersymmetry  the  background fields are then  not
modified by $\a'$ corrections \Jack\bsfet
\ref\BSslsu{I. Bars and K. Sfetsos, \pl B301(1993)183.}
\aat.

The key observation made below is that it is possible to construct
 closely related  but  non-trivial
(1,0) supersymmetric model  by giving up the $2d$ gauge  invariance already
at the classical level
and  replacing the   gauged WZNW model by the  chiral gauged WZNW one,
i.e. by using the (1,0) supersymmetric CWZNW as a starting point for
a construction of a
heterotic  $\s$-model which has an exact  conformal field
theory interpretation.
Practically,
this means  setting to zero the coefficient of the $A\A$-type term in the GWZNW
action
or, equivalently, adding the anomalous gauge degree of freedom  to the set
of  dynamical variables  and thus
remaining on the target space of  dimension equal to $\dim G$ (and not to $\dim
G/H$).
Then  no
inconsistency appears at the quantum level.  In contrast to the
GWZNW case,   the world-sheet theory  here has  only (1,0)
(and not (1,1))  supersymmetry, the target
space dimension of the corresponding heterotic $\s$-model  is  $D=\dim G$
and the background metric  receives  $1/k $ (or $\a'$) correction.
The first possibility for a non-trivial solution is  thus  in $D=3$,
corresponding, e.g.,  to $G=SL(2,\IR)$ and $H=\IR$. From the point of view
of the heterotic string theory, this $D=3$ model is the simplest yet  basic
example, similar to what   the $SL(2,\IR)/\IR$ $D=2$ `black hole' is for the
bosonic or (1,1)
supersymmetric string theory.

We shall start in Section 2 with a description of the path integral
formulation
of the GWZNW and CWZNW models  clarifying   their close   connection.
We shall establish their  formal equivalence relation (eq.(2.22)) and
illustrate
it by computing the central charges.
In Section 3 we shall  compare  the expressions
for the Hamiltonians of the  corresponding conformal field theories
demonstrating
that  the CWZNW Hamiltonian is different from the combination of the  standard
Hamiltonians of the `left' ($G/\HL$)  and `right' ($G/\HR$)  coset models.
In  Section 4 we shall consider the case when  the subgroup $H$ is abelian
and   identify $G/H$ CWZNW model with a particular axially gauged $(G\times H)
/H$
WZNW model.

The (1,1) and (1,0) supersymmetric   versions of CWZNW theory will be
discussed in Sections 5 and 6.
We shall first review the  manifestly supersymmetric
path integral quantisation of the  (1,1)  GWZNW   and  then apply a similar
approach to the CWZNW
case (discussing also the component formulation).
In Section 6 we shall treat the  case  of  (1,0) supersymmetric  CWZNW model,
establishing, in particular,  the expression for its  effective action
which,   in contrast to the (1,1) supersymmetric  case,  will contain a quantum
correction  term.
Finally, in Section 7  we  shall apply the effective action approach
\tsw\bs\aat\sfet\ to derive  the exact
(in $1/k$) form of the background fields of the
corresponding heterotic string solutions and present the first non-trivial
$D=3$ example which is
 similar  to  the  $SL(2,\IR) \times \IR /\IR$  GWZNW model.
Section 8 will contain some concluding remarks.

\newsec {  Gauged  and `chiral gauged' WZNW models }

\subsec {Path integral}
\noindent
Consider the classical action of the form
\eqn\action{I_a(g,A) =  I_0(g,A) + { a\ov \pi}  \int d^2 z \Tr( A\A) \  ,}
where  $A=A_z, \ \A= A_{\bar z}, \  $
$a$ is a constant parameter and $I_0(g,A) $ is the   gauged
WZNW action
  \witt\GWZNW\
\eqn\GWZNW{\eqalign
{I_{GWZNW}=&I_0(g,A) = I(g)  +{1\over \pi }
 \int d^2 z \Tr \bigl( A\,\bd g g{\inv} -
 \bar A \,g{\inv}\del g + g{\inv} A g \bar A  - A \A \bigr) \ ,  \cr
&I(g) ={1\over 2\pi}\int d^2 z
\Tr(\del g{\inv}\bd g)+{i\over 12\pi} \int \Tr(g{\inv}dg)^3 \ . \cr}  }
\no
The action \GWZNW\ is invariant under  the  standard  vector
gauge transformations ($A, \A$ take values in the algebra $\L(H)$  of the
subgroup $H$)
\eqn\tran{ g \to u{\inv} g u \ , \  \ A \to u{\inv} ( A - \del ) u \ , \ \
 \A \to u{\inv} ( \A - \bd ) u \ , \ \  \ \ u = u (z, \bar z) \in H \ .}
\no
For $a=1$ the action (2.1) is the CWZNW action of \gin\tye
\eqn\CWZNW{I_{CWZNW} = I_1(g,A)=I(g)  +{1\over \pi }
 \int d^2 z \Tr \bigl(  A\,\bd g g{\inv} -
 \bar A \,g{\inv}\del g + g{\inv} A g \bar A \bigr)\ .}
\no
This action is invariant under the following gauge-type transformations
\eqn\tranc{\eqalign
{& g\to u{\inv} g \bar u\ ,\ \ A\to u{\inv}(A-\del)u\ ,\ \
 \A\to \bar u{\inv}(\A-\bd)\bar u \ , \cr
& u=u(z) \in H\ , \ \ \bar u=\bar u(\bar z)\in H\ .\cr} }
 \no
In general one can consider the `left' and `right' subgroups
 of $G$ to be different, i.e. $u\in H_R$, $\bar u \in H_L$,  $A\in \L(H_R)$ and
$\A\in \L(H_L)$ with
$H_R\ne H_L$.\foot {In our notation the `left' and `right' in the group action
sense and the world
sheet $(z,\bar z)$ sense  are `cross-related' in the action (but not in the
Hamiltonian), i.e. $H_L$
acts from the left not on $g$ but on $g\inv$.  }
Since the chiral gauge transformations do not actually eliminate
dynamical degrees of freedom (since  $u $ and $\bar u$
are holomorphic  and antiholomorphic  functions)
it is more appropriate to consider them as global symmetry
transformations of the action  \action\ with $a=1$.

Parametrising  $A$ and $\A$ in terms of $h$ and $\bh$  which
take values in  $H$
\eqn\tranh{ A = \del h h{\inv} \ \ , \ \ \ \A =
\bd \bh \bh{\inv}  \ \ , }
\no
one can use  the  Polyakov-Wiegmann identity \polwig\  to represent the
action  \action\  in terms of one  WZNW action
corresponding to the group $G$
and  two WZNW actions  corresponding to the  subgroup $H$,
\eqn\acth{\eqalign{
&I_a(g,A) = I({\tilde g} ) - I({\tilde h} )
+  a\  [\ I({\tilde h} ) - I(h{\inv}) - I(\bh)\ ] \ , \cr
& {\tilde g} = h{\inv} g \bh \ , \ \ \ \  {\tilde h}  = h{\inv}\bh\ .\cr } }
\no
Clearly,  the action $I_a$  \action\ (or equivalently \acth)
is {\it classically} conformally invariant for any value of $a$.
However, it is only  for  $a=0$ or $a=1$ that $I_a$ reduces to a sum of
WZNW actions for {\it independent}  fields and  only in these
two cases it is obvious that  conformal invariance is preserved at the quantum
level.
\foot {Let us note that in the acse of the abelian $H$ the  action (2.1)
can be interpreted as a gauge-fixed form of the action of the $G\times H/H$
GWZNW nodel with $a$ being related to the parameters of embedding of $H$ into
$G\times H$ (see eq. (4.5)). Being understood in this way,  the action
(2.1) corresponds to a conformal theory for all $a$. In particular, the  \sm
obtained by integrating out $A,\A$ should also be conformal invariant
for any $a$. An example of such theory is provided by the
 $SL(2,\IR) \times \IR /\IR$  GWZNW model (the  conformal invariance of the
associated \sm  was checked  at the two-loop level in \sfts).}
Namely, we find
\eqn\ao{
a=0 :  \  \ \  \ \ I_0=I_{GWZNW}= I({\tilde g} ) - I({\tilde h} )\ ,}
\no
and for
\eqn\ae{
a=1 :  \ \   \ \ \  I_1=I_{CWZNW}= I({\tilde g} ) - I(h{\inv}) - I(\bh) \ .}
\no
For the same values of $a$ an extra local or `semi-local' symmetry appears
in \acth\ which corresponds to the gauge symmetries \tran, \tranc\ :
\eqn\extra{\eqalign{
&a=0 :
\ \ \ g\to u{\inv} g u\ ,\ \ h\to u{\inv} h\ ,\ \  \bh\to u{\inv} \bh\ , \ \
u=u(z,\bar z) \in H \cr
&a=1 :
\ \ \ g\to u{\inv} g \bar u\ ,\ \ h\to u{\inv} h\ ,\ \ \bh\to \bar u{\inv} \bh\
,
\ \ u(z), {\bar u}(\bar z) \in H\  . \cr }  }
\no
For all  other values of $a$ the above symmetries degenerate to global ones
with constant transformation parameters.
It is clear that \ae\ (but not \ao)
admits a straightforward generalisation to the case
when $h$ and $\bh$ belong to diffent subgroups $H_R$ and $H_L$ of $G$.

The corresponding path integral has the form
\eqn\Zp{\eqalign{
Z_p &= \int [dg][dA][d\A ]   \exp [- kI_a (g,A)] \cr
&= J_0 \int [dg] [dh] [d\bh ] \exp \{- kI({\tilde g} ) + (k+ 2 {\rm g_H})
I({\tilde h} )
- p [ I({\tilde h} ) - I(h{\inv}) - I(\bh)] \} \ ,} }
\no
where $p\equiv  ak - 2 q {\rm g_H}$ and ${\rm g_H}$ is the dual Coxeter number
for the
subgroup $H$. We have used the fact that  there exists  a freedom of
introducing  a local counterterm  $\sim q \Tr(A\A)$
in the definition of the Jacobian
\eqn\Jac{\eqalign{
&J= \det D(A) \det { \bar D }(\A)  = J_0   \exp \{2 {\rm g_H} [\ I(h{\inv}\bh)
+
{  q\ov \pi}  \int d^2 z \Tr (A\A)\ ]\} \  , \cr
& J_0= [\det \del\bd\ ]^{d_H}   \ ,\cr }  }
\no
of the
transformation  from $A,\A$ to $h,\bh$.  In the GWZNW case $q$  is chosen to
be zero in
order to preserve the vector gauge symmetry \karabali.
In the CWZNW case the natural choice
is the `left-right decoupled' one $q= -1$,
i.e.   $J\sim \exp \{ 2 {\rm g_H} [ I(h{\inv}) + I(\bh) ] \}$ \tye.
It is only  for the two values of $p$
\eqn\Zo{\eqalign{
p=0:   \ \ \ \ \ &Z_0=Z_{GWZNW} = J_0 \int [d{\tilde g} ][d{\tilde h} ]
\exp [ - S_{GWZNW} ({\tilde g} ,{\tilde h} ) ] \ , \cr
& S_{GWZNW} = kI({\tilde g} ) - (k+ 2 {\rm g_H}) I({\tilde h} ) \ , \cr} }
\no
and
\eqn\Ze{\eqalign{
p= k+2 {\rm g_H} :\ \ \  \ \ &Z_1=Z_{CWZNW}  =  J_0 \int [d{\tilde g} ]
[dh][d\bh]
 \exp [- S_{CWZNW} ({\tilde g} ,h, \bh)]  \ ,\cr
&S_{CWZNW}=
kI({\tilde g} ) - (k+ 2 {\rm g_H}) [ I(h{\inv}) + I(\bh)] \  ,\cr } }
\no
that the resulting quantum theory reduces  to a combination of independent
WZNW  theories and therefore  is guaranteed to be conformally invariant.
In fact, a formal proof of conformal invariance  for  the theories \Zo\ and
\Ze\ reduces to that for the WZNW model and an observation that  conformal
invariance (or UV finiteness) property is  essentially preserved under field
redefinitions.

The  quantum effective  actions corresponding to GWZNW \GWZNW\  and CWZNW
\CWZNW\ models are obtained
(up to a non-local field redefinition \aat\ which we shall ignore) by replacing
$k$  and $-k$ by $\kg$ and $-k +  {\rm g_H} $ in \Zo\ and \Ze\
(or  by multiplying the
$G$ and $H$ terms in \ao\ and \ae\  by $\kg$ and $\kh$) \tsw\bs\aat\sfet
\eqn\GGWZNW{\eqalign{
\G_{GWZNW} (g,A) &= (\kg )I(h{\inv} g \bh) -  (\kh )I(h{\inv} \bh) \cr
    &= (k + {\rm g_G}) I_{GWZNW} (g, A) + ( {\rm g_G} -   {\rm g_H})  \O (A) \
, } }
\no
and
\eqn\GCWZNW{\eqalign{
\G_{CWZNW} (g,A) &= (\kg )I(h{\inv} g \bh) -  (\kh ) [ I(h{\inv})  +  I (
\bh)]\cr
 & = (k + {\rm g_G}) I_{CWZNW} (g, A) + ( {\rm g_G} - {\rm g_H}) [ \o (A) + \W
(\A)]\ ,\cr} }
\no
where $\O (A)$  is a non-local  gauge invariant functional of $A $ and $ \A $,
\eqn\Om{
\O (A) \equiv I(h{\inv} \bh) = \o (A) + \W (\A) + \1p \int d^2 z \Tr (A\A)\ ,}
\no
and the functionals $\o$ and $\W$ are given by
\eqn\om{\eqalign{
& \o (A)  \equiv  I(h{\inv}) = - \1p \int d^2z
\Tr \{ \half A{\bd\over \del} A  + \third A[{1\over \del}A , {\bd\over \del}A]
+  O(A^4)  \}\ ,\cr
& \W(\A)  \equiv   I(\bh)
= -\1p \int d^2z \Tr \{\half \A{\del\over \bd} \A  -  \third \A[{1\over \bd}\A,
{\del\over \bd}\A] +
O(\A^4) \} \  .\cr } }
\no
As at the classical level, the effective action of the CWZNW is obtained
from the effective action of GWZNW by dropping out the $A\A$-terms
(which, of course, were    crucial  for gauge invariance of \GWZNW\ and
\GGWZNW).

Let us note that in the  general case of $H_L \not= H_R$   when $h$ and $\bh$
belong to $H_R$ and $H_L$  the action $S_{CWZNW}$ in \Ze\ and $J_0$ in \Jac\
are  replaced by
\eqn\Slr{S_{CWZNW}= kI({\tilde g} ) - (k+ 2 {\rm g_\HR}) I(h{\inv})
- (k+ 2 {\rm g_\HL})I(\bh) \  ,}
\no
and
\eqn\Jlr{J_{0R}J_{0L}= (\det \del)^{d_\HR} (\det \bar \del)^{d_\HL} \ .  }
\no
Eq. \Jlr\ implies that if $\dim H_R \not= \dim H_L$ the numbers of the `left'
and `right' bosonic degrees of freedom do not match and the theory is
 thus really `chiral' in the  $2d$ sense,
i.e. it has a Lorentz anomaly on a curved $2d$ background.
Using it to construct a consistent string theory,   one needs to compensate
the anomaly by introducing  extra chiral degrees of freedom.


 \subsec {Relation between  `chiral gauged' and gauged   WZNW models}
\no
The models \ao ,\ae\ or \Zo ,\Ze\  are particular representatives  of a
general class of models which can be called `twisted' products of WZNW models,
\eqn\twist{S= \sum_{i=1}^N \k_i I_{G_i} ({\tilde g} _i) \ ,}
\no
where $I_{G_i}$ stands for a  WZNW action for a group $G_i$;
 the arguments are related to some `original' variables  $g_i$
by field redefinitions respecting global symmetries
(e.g. ${\tilde g} _i$  are  given by particular  products of some  $ g_i$).
Such models  are  conformally invariant  and  unless there are some
`accidental' gauge symmetries (as in the case of GWZNW)
 they can be represented as $\s$-models on the target space (of dimension
$\sum_{i=1}^N \dim G_i$) equivalent to the direct product of the group spaces.
For example, in the CWZNW case \Ze\ we can parametrise  $g,
h,  \bh $  in terms of local group coordinates $X_g, X_h, X_\bh$;
then the coordinates corresponding
to ${\tilde g} $ in \acth\ (${\tilde g} = \exp (T\cdot X_{{\tilde g} })$)
are given by a local transformation
of  $X_g, X_h, X_\bh$. As a result, the action \Ze\ will  take the form
of a $\s$-model  on the group space $G\times H_L\times H_R$  represented in
terms of `transformed' coordinates.

All becomes less trivial once we  decide to treat some  subsets of the fields
among $g_i$ as more
`fundamental' than others (by constructing observables in terms of them only,
i.e. by introducing sources in the path integral  for them only).
Such a  split  may  be motivated by locality considerations: one is prompted
to
treat
$h$ and $\bh$  in \ao ,\ae\  as  auxiliary fields (which should be integrated
out first)
by the observation that being expressed in terms of the corresponding currents
$A, \A$ in \tranh\ the classical
actions \GWZNW , \CWZNW\  are local
(while they are non-local being expressed in terms of the current
corresponding to $g$).
Another possible reason may  be the elimination of negative norm fields
associated  with the negative sign terms in \twist\  (as in the case of the
GWZNW models).

 Once  one first integrates  over a
subset of  `auxiliary' fields
($2d$ gauge fields in the case of  GWZNW and CWZNW models)
one induces (on a curved $2d$ background)
the dilaton term in the  effective action. The dilaton term   is
necessary  in order to
preserve  the Weyl invariance of the resulting lower dimensional
$\s$-model \wit.  Thus the
appearance of the dilaton  can be considered as  an artifact of  concentrating
on an `intermediate'
(or `reduced') theory with  a smaller  number of fields than the original one.

If one  does not  make a separation into     `fundamental' and `auxiliary'
fields  one  may be able
to establish  various  formal equivalences between the models in \acth.
For example, as we  will  show below, the $G/H$ chiral gauged WZNW model can
be represented
as the  gauged WZNW model for the coset $(G_k/H_k) \times  H_{-k-2 {\rm g_H}}$
in the
sense that the corresponding quantum actions in \Zo\ and \Ze\ are related by a
field redefinition.
This is not too surprising  since at the level of the  path integral \Zo, \Ze\
the two theories   are represented by combinations of WZNW  models that can be
directly related.

To  demonstrate  the  formal equivalence
\eqn\euqiv{\eqalign{
(G_k/H_k)_{CWZNW}  &=   [ (G_k/H_k) \times  H_{-k-2 {\rm g_H}} ]_{ GWZNW}\cr
&= [ G_k\times H_{-k-2 {\rm g_H}} \times  H_{-k- 2 {\rm g_H}}]_{WZNW} \  ,\cr }
}
\no
let us start with  the  quantum effective CWZNW action \GCWZNW\
and represent it as a sum  of the  quantum actions for
the $G_k/H_k$ GWZNW and $H_{-k- 2 {\rm g_H}}$ WZNW by making  a  field
redefinition
\eqn\eqiv{\eqalign{
\G_{CWZNW}(G/H) &= (\kg) I(h{\inv} g\bh ) - (\kh) [ I(h{\inv}) + I(\bh)]  \cr
&= [\ (\kg) I(u{\inv} f \bar u) - (\kh) I(u{\inv}\bar u)\ ] - (\kh) I(v) \cr
&= \G_{GWZNW}(G_k/H_k) + \G_{WZNW}(H_{-k-2 {\rm g_H}}) \ ,\cr } }
\no
where $f\in G$, $u,v\in H$ and
\eqn\tr{g= {\bar u}{\inv} f \bar u v{\inv}\ , \ \ h= {\bar u}{\inv}  u\ ,
\ \ \bh = v \ . }
\no
The same equivalence relation is true also between  the classical actions in
(2.2) and (2.4)
and the `quantum' actions in (2.13) and (2.14).
The fields of CWZNW are obviously invariant under the gauge transformations of
the fields
($f, u,\bar u$) of  $G/H$ GWZNW.     Since transformation
\tr\  mixes different fields the `intermediate' $\s$-models obtained
by integrating out the gauge fields,  in the GWZNW and CWZNW models  will,  in
general,  be
different.\foot {For example, the $SL(2,\IR)/\IR \times \IR$ GWZNW theory
reduces to  the
`uncharged black string' $D=3$ $\s$-model \horne\ while
$SL(2,\IR)/\IR$ CWZNW  theory gives a `charged black string'
$D=3$ $\s$-model \kumar\sfet.}

In the case when $H$ is abelian  it is possible to establish  a more explicit
 equivalence  between the $G/H$ GWZNW and axially gauged
$(G\times H )/H $ WZNW models with a specific embedding of $H$ (see  Sect.4).
In this case the gauge fields are not transformed  and thus
the corresponding  $\s$-models (with $G$ as a configuration space) are
equivalent.


In contrast to $G/G $ GWZNW model which is topological and therefore trivial
from the $\s$-model pont of view, the $G/G $ CWZNW model is non-trivial
and equivalent to  $G_{-k-2 {\rm g_G}}$ WZNW model.
In order to prove that one uses the expression for the $\s$-model obtained
from a general CWZNW model after integrating out the gauge fields \sfet.
One can show that if $H=G$ it takes the form of the $\s$-model appropriate
for the $G_{-k-2 {\rm g_G}}$ WZNW theory.
This conclusion is of course  in agreement with \euqiv.


The  equivalence \euqiv\ implies that  if  $G$ and $ H$
are  compact the resulting  CWZNW  model will
contain $\dim H$ negative norm fields (or `times') and therefore the
physically interesting case is
only that of an abelian $H$ with  $\dim H=1$
(the models considered in \kumar\maha\sfet\ belonged to this class).

\subsec { Central charge }
\no
The  formal  relation  \euqiv\  between  the path integrals of  GWZNW and
CWZNW  models implies the equality of the corresponding central charges.
 Adding the quantum shifts to the levels,  $kI(g) \ra (\kg)I(g) , \ kI(h) \ra
(\kh)I(h)$ and accounting for the $J_0$ contribution in \Jac\
it is straightforward to compute the
values of the central charges for GWZNW \Zo\ \karabali\ and  CWZNW \Ze\
models ($d_G=\dim G, \ d_H=\dim H$):
\eqn\centg{C_{GWZNW}(G/H)  =  {kd_G\ov k + {\rm g_G}}  +
 {(-k-2 {\rm g_H})d_H\ov  (-k-2 {\rm g_H} ) + {\rm g_H}}  - 2d_H
= {kd_G\ov k + {\rm g_G}}  - {kd_H\ov k + {\rm g_H}} \equiv C_{G/H} \  ,}
\no
and
\eqn\centc{C_{CWZNW} (G/H) = {kd_G\ov k + {\rm g_G}}  +
 2{(-k-2 {\rm g_H})d_H\ov ( -k-2 {\rm g_H}) + {\rm g_H}}  - 2d_H
= C_{G/H}  + {(-k-2 {\rm g_H})d_H\ov  (-k-2 {\rm g_H}) + {\rm g_H}}\  .}
\no
As a result, the    $G/H$ CWZNW  model
has the same central charge as the  $(G_k/H_k) \times  H_{-k-2 {\rm g_H}}$
GWZNW model
(and, in particular, it cannot be equivalent to  $G/H$ GWZNW model, cf. \tye).
If $H$ is abelian, ${\rm g_H}=0 $ and \centc\ is equal to the central charge of
the $G_k$ WZNW model.

In the case when $h$ and $\bh$ belong to different subgroups $H_R$ and $H_L$
of $G$  one finds from \Ze, \Jlr, \Slr\
the following expression for the central charge of  CWZNW
\foot { This is the expression for the
Weyl anomaly coefficient, i.e. for the `left' plus `right' central charge
$C=\ha (C_l + C_r)$.
As we have
mentioned  above,  the theory contains also the Lorentz anomaly,
proportional to  $\ha ( C_l - C_r )=
d_{ H_R} - d_{ H_L}$ . }
\eqn\centlr{C_{CWZNW} =  {kd_G\ov k + {\rm g_G}}  +
 {(-k-2 {\rm g_{H_L}})d_{H_L}\ov ( -k-2 {\rm g_{H_L}} ) + {\rm g_{H_L}} }
+ {(-k-2 {\rm g_{H_R}}) d_{H_R}\ov ( -k-2 {\rm g_{H_R}} ) + {\rm g_{H_R} } }
- d_{H_L}- d_{H_R} \ .}
\no
This expression is equal to the central charge of the
 $ G_k \times (H_L)_{-k-2 {\rm g_{H_L}}}
\times (H_R)_{-k-2 {\rm g_{H_R} }}\times \IR^{- d_{H_L}- d_{H_R}}$ WZNW model
(or  of the
$(G/H_L)_k \times (H_R)_{-k-2 {\rm g_{H_R} }} \times \IR^{d_{H_L} -d_{H_R}}$
GWZNW model  or of the
$(G/H_R)_k \times (H_L)_{-k-2 {\rm g_{H_L} }} \times \IR^{d_{H_R} -d_{H_L}}$
GWZNW model) but  is
{ \it  different}  from the central charge of the direct  product
of the `left' and `right'  parts of the $G/H_R $  and $ G/H_L$
GWZNW models (cf.\tye).
In the latter case the  central charge is
\eqn\centt{ C= \ha C_{G/H_L} + \ha C_{G/H_R}
=
{kd_G\ov k + {\rm g_G}}  - \ha {kd_{H_L} \ov k +{\rm g_{H_L}}}
 - \ha {kd_{H_R} \ov k + {\rm g_{H_R}}} \ .}
\no
Note that extra `one-halves'  in front of the $H_L$ and $H_R$ contributions
do not
actually  appear in CWZNW (the ${\rm g_H}$ contributions from the Jacobians
appear
in \Ze\ with the same coefficient as in GWZNW).
As we shall see in the next section, the
Hamiltonian of the GWZNW model is also  {\it different} from the naive
combination
of the left part of the Hamiltonian of the $G/H_L$  with  the right part of the
Hamiltonian of the $G/H_R$ GWZNW models.

\newsec {Hamiltonians }

The form of the stress tensor  for  the  GWZNW and CWZNW models
can be  easily read off
from  \Zo, \Ze\ by replacing  each  WZNW action by the   bilinear  products
of the corresponding
currents
(since the WZ-term does not depend on the  world-sheet metric).
 The zero mode of the $T_{00}$-component of the  classical stress
tensor is  the Hamiltonian $\H=L_0+{\bar L}_0$   associated with the classical
 action \ao\ or \ae.
The zero mode of the quantum stress tensor is the Hamiltonian  associated with
 the quantum effective
action \GGWZNW\ or \GCWZNW\ with shifted $k$'s. The  `$J^2$' structure of
the Hamiltonian follows from the
 similar structure of the action in the $1d$ dimensional reduction limit
(in which the WZ-term  in the action does not contribute).
For example,
\eqn\S{S= k_G I_G - k_{H_L}I_\HL - k_\HR
I_\HR \  \ra \  \H =  {1 \ov k_G}  J^2_G
- {1 \ov k_\HL} J^2_\HL - {1 \ov k_\HR }\J^2_\HR \ .}
 \no
It is instructive to derive the general expression for the Hamiltonian
corresponding to
the theory \action\ supplemented by extra  $A^2$ and $\A^2$ terms.
Such terms are not Lorentz invariant but  they  can be  considered as
originating  (in the $d=1$ reduction  limit relevant for
the derivation of the zero mode  Hamiltonian)
 from the  Lorentz
invariant  non-local  terms $A(\bd/ \del)A+ \dots $ and
$\A (\del/ \bd) \A + \dots  $  which
 appear in  the quantum effective action of the GWZNW  and CWZNW models  (see
\GGWZNW, \GCWZNW, (7.3)).

Let us start with the following action
\eqn\Sa{I =I_0(g,A) + { 1\ov \pi}
\int d^2 z \Tr ( a  A\A  + \ha \b A^2 + \ha \bab \A^2 )  \  ,}
\no
where $I_0(g,A)$ is the GWZNW action \GWZNW.
When $ a=1 , n=m=0 $   this is  the  classical CWZNW action
\CWZNW.
The case of
\eqn\cao{ a= -b \ , \ \ \  \  b  = \bab =
-  {{\rm g_G}-{\rm g_H} \ov k+{\rm g_G} }\  ,}
\no
corresponds to   the quantum  GWZNW effective  action \tsw\bs,
whereas the case of
\eqn\cat{a=1,  \ \ \  b=\bab = -  {{\rm g_G}-{\rm g_H} \ov k+{\rm g_G} }\  , }
\no
corresponds to the quantum  CWZNW effective action \sfet.
Let us find the classical Hamiltonian $\H$ of the model \Sa\ treating
$A,\A$ as non-dynamical fields that should be eliminated at the end.
One  should express  $\H$  in terms of  currents
that  should satisfy    proper Poisson  bracket current algebra  relations
 as in  the WZNW model (see e.g. \bow).
These are  the left ($g{\inv} \del g $)  and right
($-\bd g g{\inv} $) currents
expressed in terms of momenta   (which  will  now contain $A,\A$ so
that the currents  will also  depend on  $ A,\A$).
Let  $J_G, \J_G$   and $J_H, \J_H$
denote  the currents corresponding to $G$ and $H$.
In what follows we shall switch to the component
notation:  $J_G=(J^A_G),\  J_H=(J^a_H),  \ A^2= A^aA_a = - \Tr(A^2)$, etc.
%
Then $\H$  will be given  by (we rescale $\H$ by  the factor of $2$ and omit
obvious factors of
$1/(k+{\rm g_G})$ and $1/\pi$ )
\eqn\ham{\H=   \ha  J^2_G + \ha \bar J^2_G -2 J_H \A - 2 \J_H A
+ 2 (a-1)  A\A  + (b+1) A^2 +  (\bab +1) \A^2 \ .  }
\no
For $a=b=\bab =0$ one obtains the Hamiltonian for the classical WZNW action
\GWZNW\
\bow
\eqn\hamo{\H_0 = \ha J^2_G + \ha \bar J^2_G -2 J_H \A - 2 \J_H A
+  ( A - \A)^2  \ . }
\no
In contrast to \hamo\ where the $A^2$-term is singular
(as a consequence of  the gauge invariance
of the GWZNW action \GWZNW) it is straightforward to eliminate $A,\A$ from
\ham.  The singular case
\hamo\ can then be defined  as the  limit  $a,b, \bab  \ra 0$
(in this way one is able to avoid the
complications  (due to constraints)  dealt with in \bow).
We find (using that $\bar J^2_G=J^2_G$)
\eqn\hame{\eqalign{
&\H =  J^2_G
+ {1\ov \Delta}[\ -  2 (a-1) J_H\J_H + (b+1) J^2_H +  (\bab +1) \J^2_H\ ]\ ,
\cr
& \Delta\equiv (a-1)^2 - (b+1)(\bab +1)  \ .\cr} }
%
\no
The `determinant' $\Delta$ is singular  both in the classical ($a=b=\bab=0$)
and
quantum  $ (-a= b= \bab) $ GWZNW cases.
To reproduce  the  well-known expression for the  GWZNW Hamiltonian from
\hame\  let us consider the
following limit: $b=\bab , \ a= - b + \epsilon , \
\Delta=  -2(b+1)\epsilon +  O(\epsilon^2), \ \epsilon \ra 0$.
Then
\eqn\haml{\H_{GWZNW} = J^2_G + {1\ov b+1}  J_H\J_H
-  {1\ov 2 \epsilon } (J_H+\J_H)^2\ .}
\no
To get a non-singular  Hamiltonian  one is to restrict  it to  a   gauge
invariant subspace on which $J_H+\J_H=0$. Then \haml\
reduces to the standard expression\foot
{The singular term in \haml\ can be thought of as giving rise to a
$\d$-function in the path integral
approach.}
\eqn\hamf{\H_{GWZNW}= J^2_G - {1\ov b+1} J_H^2 = J^2_G - {\kg\ov \kh} J_H^2\ .}
\no
The CWZNW Hamiltonian ($a=1,  \  b = \bab$, $\Delta= - (b+1)^2$)  is given by
\eqn\hamc{\eqalign{
\H_{CWZNW}&= J^2_G - {\kg\ov \kh  } ( J^2_H +  \J^2_H )  \cr
&= \ha [\ J^2_G  -  2\ {\kg\ov \kh  }  J^2_H\ ]
   + \ha [\ \J^2_G  -  2\ {\kg\ov \kh  }  \J^2_H\ ] \ .\cr} }
\no
In contrast to the GWZNW case \hamf,
here there is no need to restrict to a sector in which $J_H +\J_H=0$.
One can show \aat\sfet\
that the $\s$-model expressions for the metric and dilaton
obtained using the quantum effective actions \GGWZNW, \GCWZNW\ coincide with
the corresponding expressions obtained in the operator approach using the
Hamiltonians \haml, \hamc.

   As we have  noted already, the structure of $\H_{CWZNW}$
directly reflects the structure
of the action  in \ae\ and \Ze. According to the relation \euqiv\
it is possible to identify the CWZNW theory
with  the  GWZNW theory for the product $(G/H) \times H$  under a proper
definition  of the currents.
Namely, it should be possible to  introduce the new currents
$J'_G$, $\bar J'_G$, $ J'_H$, ${ \J'_H} $,$ {\tilde J}_H$, ${\bar {\tilde J}_H}
$
such that on the subspace of states  satisfying $J'_H + \J'_H=0 $ eq.  \hamc\
takes the form
\eqn\hrel{\H_{CWZNW}= [\  J'^2_G  -  {\kg \ov \kh  }   J'^2_H \ ]
-{\kg \ov \kh  } { {\tilde J}_H}^2\ ,}
\no
where we have used  the fact that $ {\tilde J}^2_H =  \bar {\tilde J^2}_H$.
Here the first  and second  terms represent the
$G_k/H_k$  and  $H_{-k-2 {\rm g_H}}$  factors in \euqiv.

The Hamiltonian \hamc\
has an obvious generalisation to the case  of the different `left' and `right'
subgroups $\HL$ and $\HR$
\eqn\hamlr{\H_{CWZNW}= \ha [\ J^2_G  -  2\ {\kg\ov k + {\rm g_\HL}  }  J^2_\HL
\ ]
  + \ha [\ \J^2_G - 2\ {\kg\ov k+ {\rm g_\HR}  }  \J^2_\HR \ ] \ .}
\no
This expression is {\it different} from the straightforward combination of the
`left' part of
the Hamiltonian of  the $G/\HL$ GWZNW  with the `right' part  of the
Hamiltonian of the $G/\HR$ GWZNW
\eqn\hami{\eqalign{
\H &= [\H_{GWZNW}(G/\HL)]_l + [\H_{GWZNW}(G/\HR)]_r \cr
&= \ha [\ J^2_G  -  {\kg\ov k + {\rm g_\HL}  }  J^2_\HL \ ]
   + \ha [\ \J^2_G  -  {\kg\ov  k+ {\rm g_\HR}  }  \J^2_\HR \ ]    \ . } }
The difference is due to the crucial coefficients 2 in front of the subgroup
current terms
in the CWZNW case.   We  already  noted  a  similar difference in  the
values of the central charges.

In the case when  $\HL$ is different from $\HR$ it is not clear how to
interpret CWZNW in terms of a particular GWZNW model.
At the same  time,  the path integral analysis  of Sect.2.1 implies that
this model  can be represented as the
$G_k\times (H_L)_{-k-2 {\rm g_\HL}} \times
(H_R)_{-k- 2 {\rm g_\HR} }$  WZNW model.
That means that
while for generic  values of $a,b,\bab$ the Hamiltonian \ham\  should  not
correspond to a conformal theory,  the CGWZW Hamiltonian \hamlr\ should.
One may question how the structure of (3.11) is consistent with the known
fact that  the only
(relevant  in the present case) solutions of the affine - Virasoro master
equation
\hal\mor\ are represented by  coset (GWZNW) models.
The answer is  again implied  by the structure of the path integral \Ze\
(with $h$ and $\bh$ now belonging to $H_R$ and $H_L$):
it should be possible to define the
holomorphic  and antiholomorphic currents  $J'_G,\ J'_{H_L}, \ J'_{H_R}$
and $\J'_G , \ \J'_{H_L},\
\J'_{H_R} $ such that in terms of them (3.12) takes the form of the Hamiltonian
of
the three WZNW models
\eqn\hamil{\eqalign{
\H_{CWZNW} &= \H_l + \H_r = {J'}_G^2 - {\kg\ov k + {\rm g_\HL} }
  {J'}_\HL^2 - {\kg\ov  k+ {\rm g_\HR}  }  {J'}_\HR^2  \cr
  &=\ha [\  { J'}_G^2 - {\kg\ov k + {\rm g_\HL} }
  {J'}_\HL^2 - {\kg\ov  k+ {\rm g_\HR}  }  {J'}_\HR^2 \ ]\cr
  &+ \ha [\  \J^{\prime 2}_G -  {\kg\ov k + {\rm g_\HL}  } \J^{\prime 2}_\HL
 -   {\kg\ov  k+ {\rm g_\HR}  }  \J^{\prime 2 }_\HR \ ]
\ . \cr } }
\no
The new currents should  directly correspond to the  redefined
 variables ${\tilde g} = h{\inv} g \bh, \ h_R= h{\inv}, \ h_L = \bh $
in terms of which the  path
integral \Ze\ factorises.

\newsec { The case of the Abelian subgroup $H$}
In this section we shall prove that the chiral gauged WZNW model  with abelian
$H_L=H_R=H$ is,  in fact,
equivalent  to a specific (axially) gauged WZNW  model $(G\times H)/H$ with a
 particular  embedding
of the subgroup $H$ into  $G\times H$.
 The basic idea is to `cancel' the $A\A$-term in the  GWZNW  action \GWZNW\
in order to make it look  like the CWZNW action \CWZNW.
This is achieved   by introducing extra $H$-degrees of
freedom coupled to $A\A$ and then gauging  them  away.
To make this idea  work one is to consider
GWZNW with an {\it axially}  gauged  abelian subgroup.
In this case the action (cf. \GWZNW, \ao)
\eqn\axial{\eqalign{
I_{GWZNW}^{axial}(g,A) &= I(g)  +{1\over \pi }
 \int d^2 z \Tr \bigl(  A\,\bd g g{\inv} -
 \bar A \,g{\inv}\del g + g{\inv} A g \bar A  + A \A \bigr)  \cr
& = I(h{\inv} g\bh) - I(h \bh) \  \ , \cr } }
\no
computed for $g=I$  does not vanish  but is proportional to the integral of
$\Tr( A\A )$.
Let  us  consider the $ (G \times  H )/H$ axially gauged WZNW theory
and  represent the elements of $G\times H$ as  block-diagonal matrices
\eqn\matr{\tg = \diag (g,h_0)\ ,  \ \
 h_0 = \diag (p_1, \dots,p_{r_H})\ ,    \  \  p_a=  \exp ( y_a T_a)\ ,}
\no
where  $g$ belongs to $G$ and $h_0$ is from
$H$.
An  element of the  abelian subgroup $H$
(which  we can  take, without lack of generality, to be
generated  by the maximal  abelian   subalgebra of the algebra of $G$)
of $G\times H$  can  be   represented in the form
\eqn\repr{\tth= \diag (h, h' )\ ,
\ \  \   h= \exp (\sum_{a=1}^{r_H} x_aT_a)\ ,  \ \
\  h'  = \diag (f_1, \dots,f_{r_H})\ ,
\ \  \  f_a=  \exp (n_a x_a T_a)  \  , }
\no
where $T_a$ are the generators of the abelian group $H$
and $n_a$ parametrise the
embedding of $H$ into $G\times H$.
Our aim will be to prove that  there exist an embedding such that
the  quantum effective action of $(G\times H)/H$ GWZNW  model in the axial
gauging (the analog of \GGWZNW)  is equal to the  quantum effective
action \GCWZNW\  of $G/H$ CWZNW model, i.e. (${\rm g_H}=0$)
\eqn\toprove{\eqalign{
\G^{axial}_{GWZNW}(G\times H/H) &= \G_{WZNW} (\tth{\inv} \tg \tilde \bh) -
\G_{WZNW} (\tth \tilde \bh) \cr
&= \G_{WZNW}(h{\inv} g\bh) - \G_{WZNW}(h{\inv} ) -  \G_{WZNW}(\bh) \cr
&= \G_{CWZNW}(G/H)\ .\cr }  }
\no
In the large $k$ limit this relation implies the equivalence
of the corresponding classical actions \axial\ and \ae\ (of course for abelian
$H$ only). The GWZNW action is invariant under the  gauge transformations:
$\tg\ra \tu\tg \tu, \
 \tth \ra \tu \tth , \  \tilde \bh \ra \tu{\inv} \tilde \bh $,
where $\tu=\tu(z,\bar z)$. Therefore we can prove \toprove\ in any gauge.
It is convenient to fix the gauge  so that
$\tg = \diag (g,1)$. Then computing the GWZNW action \toprove\
 using \repr\ we find
\eqn\prove{\eqalign{
&\G_{GWZNW} (G\times H/H) =
[\ (\kg) I(h{\inv} g\bh) + \sum_{a=1}^{r_H}k'_a I(f_a{\inv} {\bar f}_a)\ ] \cr
&- [\ kI(h\bh ) + \sum_{a=1}^{r_H} k'_aI(f_a {\bar f}_a)\ ] \cr
& =  [\ (\kg) I(h{\inv} g\bh) - kI(h{\inv} ) - kI(\bh)\ ]  -
{1\ov \pi}  \sum_{a=1}^{r_H}(2k'_a n_a^2  + k) \int d^2z A^a\A^a \ .\cr }}
\no
Here  $k$ and $k'_a$  are  the coefficients of the WZNW actions corresponding
to the
$G$ and $H$ factors in the product $G\times H$, i.e. the central extensions in
the current algebras defined in $G$ and for each factor in $H$.
 Also, $A^a= \del x^a, \ \A^a = {\bar \del }{\bar x}^a
, \ \Tr (T_aT_b) = - \delta_{ab}$ and we have used that in the  abelian case
the WZNW action contains
just the quadratic term, i.e. $I(h)=I(h{\inv})$. Choosing the embedding
parameters $n_a$  that
satisfy \eqn\cond{2k'_a n_a^2  + k =0 \ , \ \ \ \ \ \forall\ a=1,\dots ,r_H\ ,
}
\no
we get the desired equality \toprove.
Eq. (4.6) implies that if $k$ is positive,   all $k'_a$  will  be negative,
i.e.
the coordinates $y_a$ corresponding to the $H$-factor in $G\times H$
will  have  the  negative signs
of their kinetic terms in the $(G\times H)/H$ GWZNW action.
This  can be considered  to be a
consequence   of  the negative sign in front of the $H$-terms in the CWZNW
action in \toprove.

Since to prove \toprove\ we did not make any field redefinitions, the
established equivalence implies the
equivalence of the corresponding `reduced' $\s$-models obtained by
eliminating the  gauge fields
from the GWZNW and CWZNW actions.
In particular, the $D=3$ models  corresponding to  the
$[SL(2,\IR)\times \IR] /\IR $ GWZNW model
(with the particular embedding of $H=\IR$)  and $ SL(2,\IR)/\IR $ CWZNW model
are, in fact,  equivalent. This provides the general explanation
for the observations made in \kumar\maha\ (to the leading order in $1/k$
expansion)
 and in \sfet\ (exactly in the $1/k$ expansion).\foot { An  apparent
disagreement in the  exact
values of the antisymmetric tensor couplings in the two models observed in
\sfet\
can be interpreted as an artifact of the procedure of determining $B_{MN}$
from the local part of the effective action.
For the $SL(2,\IR)\times \IR /\IR$   GWZNW  model
the antisymmetric tensor coupling is given just by the semiclassical
expression \bs. The  procedure of
extracting it is not, however, completely  unambiguous
(for a detailed discussion of this issue see \sfts ).}

It should be  noted  that if   we  have
started with the effective quantum action
for $(G \times H)/H$ GWZNW  model
with a general embedding of  an abelian  $H$  and with  the {\it vector}
gauging,  the result would
be  equivalent to $(G/H) \times H$ which, however,  bears no resemblance to the
$G/H$ CWZNW model.
The reason is that under the vector gauge transformation
$\tg\to \tu{\inv} \tg \tu$, the $y_a$'s are invariant and therefore there
must exist a field transformation which maps the action for the
$(G \times H)/H $  GWZNW model  to the corresponding action for the
$ G/H \times H$ one
(this was worked out explicitly for the $SL(2,\IR) \times \IR^{d-2}/\IR$ model
in \gin\sft).

\newsec {(1,1) supersymmetric chiral gauged WZNW theory}

\subsec{ (1,1) supersymmetric GWZNW model }

\lref \fuch { J. Fuchs, \np B286(1987)455; \np B318(1989)631. }
\lref\KASU{Y. Kazama and H. Suzuki,  \np B234(1989)232;
 \pl B 216(1989)112.}
Before   a discussion of the (1,1) supersymmetric  generalisation  of  the
chiral gauged WZNW model \CWZNW\
it is useful  first  to recall the   supersymmetric version of the gauged WZNW
case.
 We shall follow the   manifestly supersymmetric  (superfield)
 approach of  \aat\ (see also \GRT) and
later compare this with the component formulation \schn
\ref\witnm{E. Witten, Nucl. Phys. {B371}(1992)191.}
in connection with the $N=1$ superconformal coset models \fuch\KASU.

 The  (1,1) supersymmetric  generalisation  of
the gauged WZNW action  \GWZNW\ is given by
\eqn\hati{\eqalign{
\hat I (\hg, \hat A) &= \hat I(\hg )  +{1\over \pi }
 \int d^2 z d^2 \t  \Tr \bigl(  \hat A\,\bar D \hg \hg{\inv} -
 {\hat {\bar A}} \,\hg{\inv} D \hg + \hg{\inv} \hat A \hg {\hat {\bar A}}
 - \hat A {\hat {\bar A }} \bigr) \cr
 &= \hat I(\tilde \hg) - \hat I(\tilde {\hat h}) \ ,\cr }}
\no
where the gauge superfields $\hat A ,\hat { \A}$  take values in the
algebra of  the subgroup $H$  and
\eqn\para{\eqalign{
 &\hat A = D {\hat h} \hat h{\inv} \ ,\qq  \ \ \hat \A =
\bar D \hat{\bh} {\hat \bh }{\inv}\ ,\cr
& \tilde \hg\equiv {\hat h}{\inv} \hg \hat{\bh}\ , \ \ \qq
\tilde {\hat h} \equiv {\hat h}{\inv} \hat \bh\ .\cr } }
\no
The quantisation of the theory can be reduced to that of the two
ungauged supersymmetric WZNW theories corresponding to the group and the
subgroup
\eqn\Zoo{
Z_{GWZNW}^{(1,1)} = \int [d\hg] [d\hat A][d\hat {\A}] \
\exp \{ - k \hat I(\hg , \hat A) \}
 = \int [d\tilde {\hg }] [d\tilde {\hat h}]  \ {\cal J}  \ \exp \{  - k I
(\tilde {\hat g} ) +   k  I (\tilde {\hat h})    \} \ \ . }
\no
Here $\cal J $ stands for  the product of Jacobians of the change of
superfield variables
from  $\hat A$ to  $\hat h$  and from $ \hat\A $ to $ \hat {\bh} $  (and
includes also a gauge
fixing factor). While in the bosonic case  the  corresponding product
(regularised in the left-right
symmetric way)  is non-trivial and leads to the shift  of the
coefficient of the
$H$-term in the action,  in the (1,1) superfield case  each of the
Jacobians is   proportional
to  a field-independent factor.\foot {This happens because    the   non-trivial
contribution of the
bosonic determinant  is   cancelled  out by   the   contribution  of the
fermionic  one (this
cancellation is similar to that of   the bosonic and fermionic contributions
to  the coefficient
$k$ of  the effective action in the ungauged  supersymmetric WZNW theory).
 In fact,  as in the bosonic case,
the Jacobian of the change of variables   $\hat A \ra \hat h $
can be expressed in terms of the  path integral with the action
$ \int d^2 z d^2 \t  U ( DV + [ \hat A , V])$ , where  $U$ and $V$ are
superfields
of opposite statistics. Rewriting  this action in  component fields and
integrating them out,
it is easy to see that this Jacobian is $\hat A$-independent.}
It is important to note that in the present case   one should not
include  an extra local
counterterm  $A\A$ (needed to preserve gauge invariance in the bosonic theory)
since here the Jacobians are trivial.

The theory  can thus be  represented  as a `product' of the two  (1,1)
supersymmetric WZNW theories for
the  groups  $G$ and $H$   with  the levels $k$ and $-k$.
Since  in the (1,1) case there  is no shift of $k$  at the quantum level
\red\schn\ the
corresponding effective action is the same as the classical one
(up to a field renormalisation) \aat.
To  see this in
detail at the component level, let us  start with \Zoo,   express it  in  the
 component
notation  and make the  chiral rotations to decouple fermions from bosons (or,
equivalently, integrate  fermions out).  Then
\eqn\Zoc{ Z_{GWZNW}^{(1,1)}  =  N' \int [d\tg][d\tth ]
   \  \exp [ - (k- {\rm g_G})I(\tg)   + ( k + {\rm g_H})I(\tth) ] \ .}
\no
 Ignoring   the free-theory  factors,  we can represent the
resulting theory as  the product  of the bosonic WZNW theories for
the groups $G$ and $H$   with levels  $( k- {\rm g_G})$ and
$( k - {\rm g_H})$ (we separate
 the shift ${\rm g_H}$  corresponding to the  bosonic change of variable).

The  above approach  can be  compared with the one based  on starting with
the component formulation of the (1,1) supersymmetric gauged  WZNW theory.
One considers for both the `left' and the `right' movers the
$N=1$ superconformal field theory based on the  coset \fuch\KASU
\eqn\suko{{G_{\hat k}\times SO(\dim(G/H))_1\ov
H_{\hat k+{\rm g_G}-{\rm g_H}}}\ ,}
\no
where the shifted level $\hat k=k-{\rm g_G}$ is a result of the redefinition
of the bosonic currents necessary to decouple the fermions \fuch\KASU.

In \schn\witnm\ a Lagrangian formulation of the above models was given. One
introduces free fermions $\psi, \bar\psi$  with values in the tangent space to
$G/H$
and couples them
minimally to the gauge fields $A, \A$.  Then the  corresponding
path integral  is given by
\eqn\susyac{\eqalign{Z_{GWZNW}^{(1,1)}
&= \int [dg][dA ][d\A] [d\psi] [d\bar\psi] \ \exp \{- {\hat k} I_0(g,A) -
{\hat k} I_0(\psi,A) \} \ ,\cr
&I_0(\psi,A)={i\over 2\pi}\int d^2 z
\Tr(\bar\psi D \bar\psi + \psi  \bar D \psi )\ ,\cr } }
\no
where  the covariant derivatives are  defined as
\eqn\covd{ D=\del - [A,\ {}\ ]\ ,\ \ \bar D=\bd -[\A,\ {}\ ]\ .}
\no
The explicit form of the infinitesimal supersymmetric transformations is
\eqn\sutr{\eqalign{&\d g=i \ep\bar\psi g + i \bar\ep g \psi\ ,\qq \cr
&\d A=\d \A=0\ ,\qq \cr}
\eqalign{&\d \psi = -\bar\ep (g\inv D g +i \psi^2)\big |_{G/H}\ , \cr
&\d \bar\psi = -\ep (\bar D g g\inv -i \bar\psi^2 )\big |_{G/H}\ ,\cr } }
\no
where the supersymmetry parameters are chiral,
i.e. $\ep=\ep(\bar z)\ , \ \bar\ep=\bar\ep(z)$.
%
Integrating over the fermions in \susyac\
and going  through the same  steps as  in  the bosonic case  we finish with
\eqn\Zf{\eqalign{
Z_{GWZNW}^{(1,1)}  =  &N' \int [d\tg][d\tth ]  \
{\exp \{ - \hat k I(\tg)  +  [ \hat k  +   ({\rm g_G} -
{\rm g_H}) +  2 {\rm g_H}] I(\tth) \} } \  ,\cr
&\tg= h{\inv} g \bh \ , \ \ \tth = h{\inv} \bh\ ,\cr } }
\no
where the contribution of the fermions is proportional to
${\rm g_G}-{\rm g_H}$. This expression
becomes equivalent to \Zoc\ once  the relation
$\hat k=k-{\rm g_G}$ is being
used. It is clear that after we perform the bosonic quantum shiftings in
the $G$ and
$H$ terms in \Zoc\ the quantum effective action one obtains is the same as the
semi-classical one (cf. \ao) \aat, and the $\s$-model receives no $1/k$
corrections \Jack\bsfet\BSslsu\aat.

\subsec{ (1,1)  supersymmetric  CWZNW model  }
\no
Let us now repeat the above analysis in the case of the  (1,1) supersymmetric
extension of the
chiral gauged WZNW  action  which is obtained by dropping  out the
$\hat A {\hat {\bar A }}$ term in \hati\
\eqn\hatic{\eqalign{
 \hat I_{CWZNW}^{(1,1)}  (\hg, \hat A) &= \hat I(\hg )  +{1\over \pi }
 \int d^2 z d^2 \t  \Tr \bigl(  \hat A\,\bar D \hg \hg{\inv} -
{\hat {\bar A}} \,\hg{\inv} D \hg + \hg{\inv} \hat A \hg {\hat {\bar A}}
\bigr) \cr
&= \hat I(\tilde \hg) - \hat I({\hat h}{\inv})-  \hat I( {\hat{\bh}})\ ,
 \ \ \ \ \ \ \  \tilde \hg\equiv {\hat h}{\inv} \hg \hat{\bh} \ .\cr }  }
\no
Instead  of \Zoo\ we now get
\eqn\get{\eqalign{ Z_{CWZNW}^{(1,1)}
&= \int [d\hg] [d\hat A][d\hat {\A}] \ \exp \{ - k I_{CWZNW}^{(1,1)} (\hg ,
\hat A) \}\cr
&=\int [d\tilde {\hg }] [d {\hat h}] [d {\hat \bh}] \ {\cal J} \
\exp \{  - k I(\tilde {\hat g} ) +
 k\hat I({\hat h}{\inv}) +   k \hat I( {\hat{\bh}})   \} \ ,\cr } }
\no
where the Jacobian is the same as in \Zoo, i.e. is trivial.
As a result, the theory can be represented as a product of the three (1,1)
supersymmetric WZNW theories for the groups $G$, $H$ and $H$ with  the levels
$k, -k $ and $-k$.
Using the (1,1) supersymmetric analog of \euqiv, i.e.
\eqn\euivs{ (G/H)^{(1,1)} \ CWZNW =
{G_{\hat k} \times SO(\dim (G/H))_1 \ov H_{\hat k+{\rm g_G}
- {\rm g_H} }}  \times H_{- \hat k_H - 2 {\rm g_H}} \times
SO(\dim H)_1 \ , }
\no
where $\hat k_H=k-{\rm g_H}$,
it  can  also be interpreted as a `product'  of a supersymmetric gauged
WZNW model and a supersymmetric WZNW model. Because of the particular
structure of the levels of the various factors it is expected,
as in the case of the (1,1) supesrymmetric WZNW model \bsfet, that the
effective quantum action will receive no $1/k$ corrections.
%
%
As in the case of (1,1) supersymmetric gauged WZNW case,
one should be able to reproduce the result \get\
using the component approach.
Since the bosonic
CWZNW action is not gauge invariant, the fermionic partners
of the  bosons $g$ here  take values
in the  algebra of  $G$ itself and not in  the tangent space
to $G/H$ as in the GWZNW case \susyac, i.e. $\psi, \bar \psi \in \L(G)$. Then
the path integral corresponding to the (1,1) supersymmetric  CWZNW model is
\eqn\susyacc{\eqalign{Z_{CWZNW}^{(1,1)}
&= \int [dg][dA ][d\A] [d\psi][d\bar \psi] \ \exp \{- {\hat k} I_1(g,A) -
{\hat k} I_0(\psi,A) \} \ ,\cr
&I_0(\psi,A)={i\over 2\pi}\int d^2 z
\Tr(\bar\psi D \bar\psi + \psi  \bar D \psi )\ ,\cr } }
\no
where the covariant derivatives were defined in \covd.
The explicit form of the infinitesimal supersymmetric transformations is
\eqn\sutrc{\eqalign{&\d g=i \ep\bar\psi g + i \bar\ep g \psi\ ,\qq \cr
&\d A=\d \A=0\ ,\qq \cr}
\eqalign{&\d \psi=-\bar\ep \bigl[
g\inv (\del g -A g) +\del \bh \bh\inv +i \psi^2 \bigr] \ ,  \cr
&\d \bar\psi = -\ep \bigl[ (\bd g + g\A)g\inv -\bd h h\inv - i \bar\psi^2
\bigr]\ ,\cr } }
\no
where the supersymmetry parameters are chiral,
i.e. $\ep=\ep(\bar z)\ , \ \bar\ep=\bar\ep(z)$.
Expressed in terms of the three superfields
${\hg } , \ {\hat h}, \ {\hat \bh}$  the  CWZNW action is manifestly
(1,1) supersymmetric.
The supersymmetry is preserved if one integrates over ${\hat h}, \ {\hat \bh}$
obtaining the effective action (or, up to non-local terms, a \sm) for $\hg$.
Integrating out  only the fermionic partners  of $h$ and $\bh$
(or of  $A$ and $\A$)
while keeping the bosonic fields $h, \bh$ one  gets the action in \susyacc.
%
%
The fields $\psi$ are the fermionic components of the `rotated'
superfield $\tilde {\hg }$ in \hatic, \get, i.e. they are the combinations
of the original fermionic partners of $g,h,\bh$.
As we see from \sutrc\ the action \susyacc\ lacks  a linearly realised
supersymmetry. The formal  supersymmetry transformation laws are non-local
being expressed in terms
of $A,\A$ (but local in terms of $h,\bh$).   They  become local  once one
integrates out  $A,\A$.

As in the GWZNW case \Zf\ the combined contribution of the integrals over
the fermionic partners of  $g$
and $A,\A$  is proportional to ${\rm g_G}-{\rm g_H}$
(the latter has the opposite
sign  since it may be thought of as originating  from the measure).
Not including (as in the bosonic CWZNW  case) the
local  $A\A$ - counterterm  in the results for  both the bosonic Jacobian
and the fermionic determinants we  find  (cf. \Zf)
\eqn\mee{Z_{CWZNW}^{(1,1)}
=  N' \int [d\tg][d h ] [d \bh ]  \  \exp \{ - \hat k I(\tg)   +
[ \hat k  +  ({\rm g_G} - {\rm g_H})  +  2 {\rm g_H}] [I(h{\inv})
+ I(\bh) ] \}  \ .  }
\no
After one uses the relation  $\hat k=k- {\rm g_G}$ this  expression becomes
equivalent to the
component form of \get\ with the fermions integrated out (cf. \Zoc)
\eqn\nee{ Z_{CWZNW}^{(1,1)}
=  N' \int [d\tg][dh ][d\bh]   \  \exp \{ - (k- {\rm g_G})I(\tg)
+ ( k +  {\rm g_H})[I(h{\inv}) + I(\bh)]  \} \  .}
\no
It is now straightforward to  write
down the expression for the
effective action in the  chiral gauged (1,1) supersymmetric WZNW theory.
Using either the representation in terms
of  the  ungauged  supersymmetric WZNW theories \get\   or
the equivalent
formulation in terms of the ungauged bosonic WZNW theories \nee\   we get the
following expression for    (the bosonic part of)  the effective action
(we omit non-local terms  originating  from  field renormalisations \aat)
\eqn\efsc{\G_{CWZNW}^{(1,1)} (g,A)
= k I (h\inv g \bh ) - k   I (h{\inv})   - k I(\bh)   \ .}
\no
As in the ungauged  supersymmetric WZNW theory,    but in  contrast  to the
result in  the
bosonic  CWZNW
theory,  here there are  no shifts in the overall coefficients  of the $G$- and
$H$- terms
in
$\G_{CWZNW}$.   The local part
of the effective action  of the  (1,1) supersymmetric CWZNW  model is equal to
the
 {\it classical} action of the bosonic   CWZNW theory \CWZNW,
\eqn\efscc{ \G_{CWZNW}^{(1,1)}  (g, A)
 = k \big[ I(g)  +{1\over \pi }  \int d^2 z \Tr \bigl(  A\,\bd g g{\inv} -
 \bar A \,g{\inv}\del g + g{\inv} A g \bar A  \bigr) \big] \   ,}
\no
i.e.  in contrast to the bosonic case \CWZNW\  it does not contain the
quantum correction  term proportional to
$ b =-{ {\rm g_G} - {\rm g_H} \ov k+{\rm g_G} }$.
As a consequence,  the exact form of  the  bosonic part of the corresponding
$\s$-model is
equivalent to the
`semiclassical' form of the \sm  in  the bosonic theory.
This result is  similar  to the one  found  for  the (1,1) CWZNW
theory,  in agreement  with the  equivalence  \euqiv\  between CWZNW and
$(G/H) \times H$ GWZNW models
which was established in the bosonic case and with  similar relation
\euivs\  in the supersymmetric one.
The quantum
Hamiltonian corresponding to the (1,1) supersymmetric CWZNW model has the
same form as the classical
bosonic one,  i.e. \hamc\ in the $k\to \infty$ limit
(we consider the  case  when the left and right  subgroups are the same)
\eqn\hamsc{\eqalign{
\H_{CWZNW}^{(1,1)}
&=   \ J^2_G  -    J^2_{H} - \J^2_{H} = \ha [\ J^2_G  -   2 J^2_{H}\ ]
+  \ha [\ \J^2_G  -   2 \J^2_{H}\ ] . \cr
 } }
\no
Note  again that this  Hamiltonian is  {\it not}
equal to the sum of the Hamiltonians of the left and right $G/H$ coset models
because of the extra coefficients 2 in front of the subgroup current terms.

\newsec {(1,0) supersymmetric chiral gauged WZNW model }
Let us now turn to  a  less trivial  (1,0) supersymmetric  theory,   which  is
to be
 related to the
heterotic string  \sm \hw.
We  are going to  repeat the previous analysis,  replacing  (1,1) superfields
by   (1,0) ones.
Part of the discussion will be similar to that in \GRT.
It was concluded in \GRT\ that since the direct (1,0) truncation of the (1,1)
supersymmetric  GWZNW theory is anomalous,
  it does not describe a consistent heterotic string  background.
Here instead we shall start with the  (1,0) supersymmetric extension  of
bosonic
(or truncation of (1,1) supersymmetric) chiral  gauged WZNW model.
 Since   we shall  give up  the gauge invariance  and will not reduce to $G/H$
already at
the classical level,  no contradiction  will   be
 found at the quantum level. As we shall see below in Section 7,
the resulting model  provides a non-trivial
example of a  heterotic string solution that is not  effectively (1,1)
supersymmetric (as were solutions  in  \GRT)
but still has a well defined  conformal field theory  counterpart.

\subsec{Component approach}
\no
To illustrate the point that the  `anomalous' contribution
of Weyl fermions  does not represent a
problem in the  CWZNW case it is instructive to start with a
heuristic discussion
in  the component approach.  Using the
component form of the action of the (1,1) supersymmetric
CWZNW  and dropping  out
the right  component ($\bar \psi$)  the fermionic fields
 we  shall  add instead some `right' internal fermions $\bar \psi^I$
which are not coupled to $A, \A$ (but may be coupled
to a background target space gauge field
 which
in the present  case we shall set equal to zero) and do not transform
under supersymmetry. Then we get
\eqn\hetact{I_{CWZNW}^{(1,0)} (g,A,\psi, \bar\psi^I)
= I_{CWZNW}(g,A) + {i\ov 2\pi}  \int d^2z \Tr (\psi \bar D \psi)
+ I_{int} (\bar \psi^I) \ .}
\no
If one
changes  the bosonic variables  from $A,\A$ to $h, \bh$
(without introducing the local
$A\A$- counterterm) and  integrates
over  the fermions  $\psi, \bar \psi^I$  one finds  the following expression
for the quantum partition
function
\eqn\hetac{\eqalign{ Z_{CWZNW}^{(1,0)}
&=   \int [d g][d h ][d\bh ]  \exp \{ -  \hat k I(h{\inv} g\bh )
+( \hat k  + 2 {\rm g_H}) [I(h{\inv})  + I(\bh) ]  +
 ({\rm g_G} - {\rm g_H}) I(\bh) \} \cr
&= \int [d \tg][d h ][d\bh ]    \  {\exp \{ - \hat k I(\tg)
+ [\hat k  +  2 {\rm g_H} + ({\rm g_G} - {\rm g_H}) ] I(\bh) + (\hat k
+  2 {\rm g_H}) I(h{\inv}) \} } . \cr } }
\no
Here the $({\rm g_G}-{\rm g_H})$-term is the fermionic contribution.
This expression is to be compared with  \mee, \nee\ found in the (1,1)
supersymmetric case. Using $\hat k=k-{\rm g_G}$
as in the (1,1)  supersymmetric case\foot {The shift of
$k$  that occurs  in the  (1,1) WZNW  model
must still be present  in the    (1,0) case   since going from superfields
to
components we  still  get the  coupling of  the Weyl fermions to the
$g$-dependent
current  and, as a result,
the anomalous contribution  that shifts the coefficient $k$
of the  bosonic term in the action.}
we find
\eqn\ff{ Z_{CWZNW}^{(1,0)}=   \int [d \tg][d h ][d\bh]
 \  \exp \{ -  (k- {\rm g_G}) I(\tg )
 +  ( k - {\rm g_G}  +  2 {\rm g_H}) I(h{\inv})   + ( k + {\rm g_H})
I(\bh)   \}  \  .}
\no
To get the  effective action
corresponding to \ff\ one is to make  further (bosonic) shifts of  the levels.
 Up to a non-local field redefinition we get (cf.\GCWZNW)
\eqn\hetef{\eqalign{
\G_{CWZNW}^{ (1,0)} (g,A) &= k I (\tg ) - (k - {\rm g_G} + {\rm g_H})
I (h{\inv})   - k I(\bh)    \cr
& = k I_{CWZNW}(g,A) +  ({\rm g_G}-{\rm g_H}) \o (A) \ ,\cr } }
\no
where $\o$ was defined in \om.
We conclude that in contrast to the (1,1) supersymmetric  model,
in the  (1,0) supersymmetric  case   there is
 a {\it non-trivial} quantum correction in the effective action (or in the
quantum Hamiltonian of the corresponding  conformal theory).
 The expression \hetef\ can be obtained from its  bosonic
counterpart \GCWZNW\ by dropping out the  $\A$-part of the quantum correction
term and replacing  the  shifted  $k$  by the unshifted one in the
classical term.

According to \ff\ the  bosonic sector of the (1,0) theory is  represented by
 a combination of  the  three
WZNW theories for $G$, $H$  and $H$.
 Comparing
(6.2) with (5.11) we can interpret this  theory  as a  `product' of the
supersymmetric  $(G/H)_k= G_{k-{\rm g_G}}/H_{k- {\rm g_H}}$  GWZNW model and
the bosonic  $H_{-k- {\rm g_H}}$  WZNW model (cf.(2.16)).
Since the  resulting theory is not (1,1) supersymmetric, it is not surprising
that here we
will have non-trivial $1/k$ corrections in the effective action (and, as a
result,  in the
corresponding target space background fields).\foot {Note
that the quantum correction in (6.4) is absent when $G=H$. In this case  the
theory
effectively reduces to the  (1,0) supersymmetric  WZNW theory.}
  We shall return to the discussion of
the effective action \hetef\ in Section 7  below.

\subsec{Superfield approach}
\no
Let us now   support   the above  component  analysis by  using   directly
 the (1,0) superfield formulation.
To obtain  a  manifestly (1,0) supersymmetric  CWZNW action one starts with
the (1,1)
supersymmetric action (5.10) and truncates the (1,1) superfields to (1,0)
superfields (in  the
heterotic string context we should  also add   the  internal (1,0) superfields
$\Psi^I$ which
 will not,   in
contrast to the case  considered in \GRT,  be important in the present
discussion). The
resulting action is
  $$ \hat I_{CWZNW}^{(1,0)} (\hg, \hat A, \Psi) = \hat  I^{(1,0)}(\hg )
+{1\over \pi }
 \int d^2 z d \t  \Tr \bigl(  \hat A\,\bar \del \hg \hg{\inv} -
 {\hat {\bar A}} \,\hg{\inv} D \hg + \hg{\inv} \hat A \hg {\hat {\bar A}}
  \bigr)\  + I_{int} \ ,
 \eq{6.5} $$
where $\hat I^{(1,0)}$ denotes the (1,0) supersymmetric WZNW action \hull
$$ \hat I^{(1,0)}(\hg)  \equiv  {1\over 2\pi }
\int d^2 z d\t  \{ \Tr (D \hg{\inv}
\bd \hg )  - {i}   \int dt [\hg{\inv} D\hg , \hg{\inv} \del_t\hg ] \hg{\inv}
\bd \hg \  \} \ ,
\eq{6.6} $$
and
$$
I_{int} = \int d^2z d\t   \Psi^I   D  \Psi^I \ , \ \ \ \
\hat g= \exp (T_A X^A) \ , \  \ \ \ X^A= x^A + { \t }{ \psi}^A_+  \ , \ \ \ \
\ \Psi^I= \psi^I_- + { \t }{ f}^I  \ \ , $$ $$  \   \hat A =  \chi_+  + \t A  =
 D\hat h  {\hat
h}{\inv} \ , \ \ \  \hat \A = \A + \t \chi_-  =\bd \hat{ \bh } {\hat \bh
}{\inv}  \  . \eq{6.7}
 $$
 We get
 the following path integral
$$ Z_{CWZNW}^{(1,0)} = \int [d\hg] [d\hat h][d\hat {\bh}]   \ {\cal J}'  \ \exp
\{  - k I^{(1,0)}
({\hat h}{\inv} \hg \hat{\bh} ) +
   k  [I^{(1,0)} ({\hat h}{\inv}) + I^{(1,0)}( \hat \bh)]    \} \ \ , \eq{6.8}
$$
where ${\cal J}'$ is  the product of Jacobians of the two changes of variables
$ \hat A \ra \hat
h $
and $ \hat \A \ra \hat \bh$.  As  discussed in \GRT,
the first Jacobian is essentially the same as in the bosonic case while the
second  one is still
trivial as in the (1,1) supersymmetric case. Therefore,
$$ Z_{CWZNW}^{(1,0)}=  \int [d\hg] [d\hat h][d\hat {\bh}]   \   \ \exp \{  - k
\hat I^{(1,0)}
({\hat h}{\inv} \hg \hat{\bh} )  $$ $$  +\
     k  [\hat I^{(1,0)} ({\hat h}{\inv} ) + \hat I^{(1,0)} ({\hat \bh})]
+ 2 {\rm g_H} \hat I^{(1,0)} ({\hat h}{\inv} )  \} \ \ . \eq{6.9} $$
There is  also an extra anomaly term originating from
non-invariance of the path integral measure  under the (1,0) superfield
rotation of $\hg $
 \GRT\ so that the final   result  is
$$ Z_{CWZNW}^{(1,0)} =  \int [d\tilde \hg] [d\hat h][d\hat {\bh}]
  \   \ \exp \{  - k \hat I^{(1,0)}
(\tilde { \hg}  )   $$ $$  +\
     k  [\hat I^{(1,0)} ({\hat h}{\inv} ) + \hat I^{(1,0)} ({\hat \bh})]
- ({\rm g_G} -{\rm g_H})
\hat I^{(1,0)} ({\hat h}{\inv} )  \} \ \ . \eq{6.10} $$
The corresponding   superfield effective action
$$ \G^{(1,0)}_{CWZNW}  = k \hat I_{CWZNW}^{(1,0)} (\hg, \hat A) +
({\rm g_G} -{\rm g_H})
\hat I^{(1,0)} ({\hat h}{\inv} )  \ \eq{6.11} $$
 is thus perfectly consistent with the expression (6.4) found
above  in the   component approach.

In the case of CWZNW model with two different subgroups ${H_L} $ and $H_R$
its  (1,0)  supersymmetric extension  is different  from the (0,1) one.\foot
{Using
 such models  for the
construction of  the heterotic string  solutions  we  need to
 compensate for a mismatch between the
numbers  of left and right  bosonic degrees of freedom  (see
the remark at  the end of Section 2.1).
}  The corresponding (bosonic parts of) Hamiltonians of  the  (1,1), (1,0)  and
(0,1)
supersymmetric CWZNW models  can be found  using the
general expressions (3.1),(3.4) (cf.(3.11),(5.16))
$$ \H_{CWZNW}^{(1,1)} =  \ha [ J^2_G  - 2 J^2_\HL]
+ \ha [ \J^2_G  -  2  \J^2_\HR]  = \ha  [ J^2_G - J^2_\HL -  \J^2_\HR ]\ ,
\eq{6.12} $$
$$\H_{CWZNW}^{(1,0)} =
 \ha [ J^2_G  - 2{  k \ov k-  ({\rm g_G}-  {\rm g_{{\HL} } })  } J^2_{{\HL} }]
+  \ha  [\J^2_G  - 2\J^2_\HR]  $$
$$ = J^2_G - {  k \ov k-   ({\rm g_G}- {\rm g_{{\HL} }})  } J^2_{\HL} -
\J^2_\HR \ ,\eq{6.13} $$
$$\H_{CWZNW}^{(0,1)} =
\ha [ J^2_G -2 J^2_\HL ]  + \ha [ \J^2_G - 2{  k \ov k- ({\rm g_G}-
{\rm g_{\HR} }) }\J^2_{\HR}] \ $$
$$ = J^2_G  -  J^2_\HL - {  k \ov k-  ({\rm g_G}
- {\rm g_{\HR} }) } \J^2_{\HR}\ . \eq{6.14} $$
\no
Note that the   Hamiltonians for the `heterotic' cases (6.13),(6.14) are  equal
to
the combinations of the left and right parts
of the Hamiltonians of the  bosonic and  supersymmetric  CWZNW models and
{\it not }
of the  Hamiltonians of the  bosonic and  supersymmetric coset $G/H$ models.
 The important difference is also that here the configuration space
is not reduced to $G/H$ but  remains the group space itself.
 In contrast to a naive  `left plus right' combination of the bosonic and
supersymmetric coset models the above models are well defined.


\newsec {Heterotic string  solutions  corresponding to  the (1,0)
supersymmetric  chiral gauged
WZNW model }
Let us now  use  the (1,0) supersymmetric  CWZNW model as a basis for a
construction  of the heterotic string solutions.
As in the bosonic or (1,1) supersymmetric GWZNW case the idea is to  treat the
$2d$ gauge field  as an `auxiliary' variable and  thus  to eliminate  it
from the  (effective) action  obtaining the \sm for the
`observable' coordinates $x^M$ of the
 configuration space ($G/H$ in the GWZNW case and group space $G$  in the
 CWZNW case).   In the case of
bosonic CWZNW model this  was already discussed
 on particular examples in \kumar\maha\ and  in general in \sfet.
 If we start  with the (1,0) supersymmetric  CWZNW theory the resulting
\sm is bound to be conformal and
(1,0) supersymmetric  and therefore to  represent a solution of the heterotic
string
theory.
\subsec {General expressions for the background fields}
\no
 As follows from the comparison of the  corresponding
effective actions \GCWZNW, \efscc\ and \hetef\ the semiclassical
(or  leading order  in $1/k$)  expressions for the  background
fields are the same in the bosonic,
(1,1) and (1,0) supersymmetric CWZNW  cases.
Like the bosonic one, the heterotic solution is modified by  the $1/k$
(or $\a'$) corrections.
The generic expression for the effective action can be represented in the form
(cf. \action, \GWZNW, \GGWZNW, \GCWZNW, \Sa, \hetef )
\eqn\general{\eqalign{ \G (g,A, \A)
=& \ \k \  [\   I(g) +{1\over \pi }
 \int d^2 z \Tr \bigl(  A\,\bd g g{\inv} -
 \bar A \,g{\inv}\del g + g{\inv} A g \bar A  - A\A \bigr)  \cr
 &+ {a \ov \pi }  \int d^2 z \Tr (A\A) -   b \o (A) -  \bab \W (\A)  \  ] \
,\cr } }
\no
where the values of the constants $a,b,\bab $  corresponding to the bosonic
GWZNW
and CWZNW models were given in \cao, \cat.
%
%
The heterotic or (1,0) CWZNW case is  intermediate between the bosonic and
(1,1) supersymmetric  CWZNW ones (in the latter case  $\k=k , \ a=1, \
b=\bab=0$)
\eqn\parahet{ \k = k \ , \ \ \ a=1\  , \ \ \ \
  b=  - {1\ov  k}   {({\rm g_G}-{\rm g_H})}\ , \ \ \ \bab =0 \ .}
\no
The derivation of the  \sm corresponding to the general  effective action
\general\ which encompasses all  known GWZNW and CWZNW cases  is given  in
\sfts.  Here for simplicity we restrict the
discussion to the heterotic case \hetef\ or \general\ with \parahet.
Dropping out the higher order non-local $O(A^3)$ terms in $\o (A)$
(since they do not affect the derivation of the local
 part of the \sm action \aat)  we obtain\foot{ This is equivalent to
approaching the point particle
limit of the theory, in which all non-local terms drop out, and then restoring
formally
the Lorentz invariance \bs.}
\eqn\hetfl{ \G^{(1,0)} (g,A, \A) =\  k \  [  I(g)  +{1\over \pi }
 \int d^2 z \Tr \bigl(  A\,\bd g g{\inv} -
 \bar A \,g{\inv}\del g + g{\inv} A g \bar A
+ \ha b  A{\bd\over \del} A \bigr) ] \ . }
\no
Introducing the notation
 ($T_A = (T_a , T_i )$ are  the
generators  of $G$;  $T_a$ are the generators of  $H$; $\ A=
1,\dots, d_G ; \ a=1,\dots, d_H; \ \eta_{AB} $ is negative definite in the
compact case)
\eqn\notation{\eqalign{
&A= A^aT_a\ , \ \ \   C_{AB} \equiv  \Tr ( T_A g T_B g{\inv} ) \ , \  \ \
 \Tr (T_AT_B) = \eta_{AB} \ ,  \cr
&J_a =  \Tr (T_a g{\inv}\del g)=  L_{aM }(x)  \del x^M \ , \ \ \
 \J_a = - \Tr (T_a \bd g g{\inv} )= R_{aM }(x)  \bd x^M \ , \cr
&{\tilde J}_a =  \Tr (T_a g{\inv}\bd g)=  L_{aM }(x)  \bd x^M \ , \ \ \
{\tilde {\J }}_a =- \Tr (T_a \del g g{\inv} )= R_{aM }(x)  \del x^M \ ,\cr
& R^A_M=-C^A{}_B L^B_M\ , \ \ \ \ C^{AD}C_{BD}= \delta^A_B \  ,\cr } }
\no
we find the following solution for $A^a, \A^a$
\eqn\sol{A=  (C^T){\inv} J\ ,\qq
\A= C{\inv} \J - b (C^TC){\inv} \tilde J + \dots
 \   ,  \qq  C=(C_{ab}) \ , }
\no
where dots stand for non-local terms.
Inserting \sol\ back into the action \hetfl\ we get for the local part of
the effective action \hetfl
\eqn\hetfll{\G_{loc}^{(1,0)}
(g,A, \A) =\  k \  \{  I(g)  +{1\over \pi }
 \int d^2 z \Tr \bigl[ - J{C{\inv}}\J  +
\ha b {J} (C^TC){\inv} {\tilde J}  \bigr] \} \ .}
\no
Identifying this action with   (the bosonic part of) the heterotic \sm action
we obtain the exact expressions  for the target space metric and the
antisymmetric tensor coupling
\eqn\hetsm{S(x)= \G_{loc}^{(1,0)}(g)=
 {k \over 2 \pi  } \int d^2 z \ {\cal G}_{MN} (x) \del
x^M \bd x^N \ ,  }
\no
where
\eqn\meax{\eqalign{
&G_{MN} \equiv  {\cal G}_{(MN )} = { G}_{0 MN }
-  2 M{\inv}_{ab}  L^a_{(M } R^b_{N)}
+  b  (M^TM){\inv}_{ab} L^a_M L^b_N\ , \cr
&B_{MN} \equiv  {\cal G}_{[MN ]} = B_{0 MN} - 2 M{\inv}_{ab}L^a_{[M }R^b_{N] }
\ , \qq M_{ab} \equiv
C_{ab} \ ,\cr }} \no
where   ${ G}_{0MN }$  and ${ B}_{0MN }$  are  the  original  WZNW
(group space) couplings,
\eqn\orig{ G_{0MN} =-\eta_{AB} L^A_M  L^B_N  \ , \qq
 3\del_{[K} B_{0MN]} =  L^A_KL^B_ML^C_N f_{ABC} \ . }
\no
We conclude that  the exact expression for the metric contains only the
leading (two-loop)
correction (recall that $b= -{({\rm g_G}-{\rm g_H})}/\k$) while the
antisymmetric tensor
is the same as in the semiclassical  approximation.
It is easy to see that the determinant  of the matrix in the
quadratic $(A,\A)$-term in \hetfl\ does not depend on $b$ and thus
the expression for the dilaton also remains semiclassical
\eqn\dil{ \p = \p_0 - \ha  {\ \rm ln \ det \ }  M   \ . }
\no
Note that the  `measure  factor'
$\sqrt G \exp(-2\p) $ still does not receive quantum
corrections  since one can prove (cf. \aat\sfet)
that $G=\det G_{MN}$ does not depend non-trivially on $b$ (there is
only an overall $b$-dependent factor).
It is also clear that when $G=H$ we get just the WZNW model with the
opposite sign in front of the first term in the action:
$G_{MN}=- G_{0 MN }$, $ B_{MN}  = B_{0 MN}$
(the dilaton in \dil\ is then constant since  according to \notation\
$|{\rm det\ } C_{AB}| =1$).

These  expressions   are to be compared with the  (1,1) supersymmetric
(superstring) CWZNW  case
where all the fields are given by semiclassical expressions
as well as with  the  bosonic CWZNW case \sfet\ where  all the fields in
general
receive quantum corrections
to all orders in the $1/k$ expansion.\foot {For completeness,
let us also recall that in the GWZNW
case  the semiclassical ($b\ra 0$) expressions for the corresponding background
fields are  formally the same  (before projection on $G/H$)
 as in \meax, \dil\ with $M_{ab} =
C_{ab} - \eta_{ab}$.
However, in this case the residual gauge invariance of the model  demands gauge
fixing, i.e. reducing the configuration space to $G/H$.  }

What is the  geometrical interpretation of the resulting spaces?
These are some  `deformations' of group spaces
with the matrix $C$ playing the role of a
deformation `parameter'.  The  deformation is related to the presence of a
non-trivial
dilaton  which,  in turn, is  necessary in order to satisfy the \sm conformal
invariance
conditions once the
model is perturbed from the  original (group space) conformal  point.
It should be stressed that the `$J^2$-perturbation' of the WZNW model in
\hetfll\ is not
in general of an integrably marginal type; the `perturbed' model
is conformal only for the specific function $C_{ab}(g)$
that appears in the CWZNW model.

\subsec{ Basic $D=3$ example: heterotic $SL(2,\IR)/\IR$  CWZNW model }
\no
Since the \sm configuration space in the CWZNW case has dimension $D=\dim G$
the first non-trivial example of the
heterotic solution based on CWZNW model   is found for $G=SU(2)$ or
$SL(2,\IR)$.
The form of  the  \sm corresponding to the bosonic $SL(2,\IR)/\IR$ CWZNW
theory was
determined in \kumar\sfet\ and it was noted that this model  is closely
related  to a specific limit  of the $[SL(2,\IR)\times \IR ]/\IR $
GWZNW theory \horne\sft\ (see Section 4).
Let us  first  recall  the  exact expressions  for the background fields in
the bosonic model.
In the bosonic case the exact
metric, antisymmetric tensor and dilaton are given
 by \sfet
\eqn\metric{\eqalign{&ds^2
= -{z-1\ov z+b }\ dt^2 - {z+1\ov z-b }\ dx^2 + {dz^2 \ov 4(z^2-1)}\ ,\cr
&B_{tx} = - (1-b){ z \ov z^2 -b^2 } \ , \cr
&\p=\p_0 - {1\ov 4} \ln (z^2-b^2)\ .\cr }  }
\no
where $b= -  {({\rm g_G}-{\rm g_H})}/\k = {2/( k-2)} $
(and $\a'= 1/(k-2)$).
\no
In the heterotic case  ($b=2/k$, $\a'= 1/k$)
  we get the semiclassical expressions
for the antisymmetric tensor and the dilaton, i.e.
\eqn\bp{B_{tx} =  -{1\ov z } \ , \ \ \qq \p=\p_0 - {1\ov 2} \ln z \ , }
\no
while  the metric  one finds  from (7.8)
\eqn\threem{\eqalign{
ds^2=&-{z-1\ov z}\ dt^2 -{z+ 1\ov z}\ dx^2 + {dz^2\ov 4(z^2-1)}   \cr
&-{b\ov 2z^2 } \bigl[  (z-1)\ dt +(z+1)\ dx \bigr]^2\  \cr } }
\no
 contains a non-trivial $O(\a')$ correction term.\foot {The variables $t,x$ in
(7.11)
have been rescalled
with respect to the original
`classical' variables of the $SL(2,\IR)$ group element,
i.e. $(t,x)\to \bigl({2/ (b+1)}\bigr)^{1/2} (t,x) $.  The corresponding
rescaling in (7.13)
 is  $(t,x)\to \sqrt 2  (t,x) $.}

It  is useful to give the explicit derivation   of
the expressions for the bosonic and heterotic $SL(2,\IR)/\IR$
backgrounds   using  the following  parametrisation
of the $SL(2,\IR)$ group element\foot {This parametrisation  was used in \dvv\
and
also in \tsw. The discussion that follows is very close to that in \tsw\
where the  case of the $SL(2,\IR)/U(1)$ GWZNW  model was considered.
 Note that in our present notation
the sign of $A$  was changed.}
$$ g= {\rm e}^{{i\over 2 } \t_L \s_2 } {\rm e}^{\ha
r\s_1} {\rm e}^{{i\over 2 } \t_R \s_2  } \ , \ \   \ \ \ \ \
  \  \t_L= \t + \tt \ , \ \ \ \t_R = \tt-\t \ . \eq{7.14}
   $$
Taking $A,\A$ to be in the  $U(1)$ subgroup  generated by $\ha \s_2$,
 the classical CWZW action (2.4)  can be represented in
 the form
$$ S_1(g,A) =kI_1(g,A) = {k \over 2 \pi } \int d^2 z [ \ \ha  (\del r \bd r  -
\del \t_L \bd \t_L
 -  \del \t_R \bd \t_R - 2 C \ \bd \t_L \del \t_R )$$
$$  - A(\bd \t_R + C \ \bd \t_L)  + \A (\del \t_L +   C  \  \del \t_R) + C A\A
\ \ ] \  , \ \ \
C=C(r)  \equiv  \cosh r \ . \eq{7.15}
  $$
The effective actions  corresponding to the bosonic and (1,0)
supersymmetric CWZNW theories  (2.16) and (6.4) (or (7.1) with (3.4) and (7.2))
 in the present model are   particular cases of  (here the subgroup is abelian
so
the quantum terms are bilinear in $A,\A$)\foot {Comparing with Section 7.1
note that  here we use a different
normalisation of the generators:
 the generator of the subgroup is $T=\ha \s_2$ so that  $\Tr A^2= \ha A^2$.}
$$ \G(g,A)  = {\k \over 2 \pi } \int d^2 z \ [ \ \ha  (\del r \bd r  -  \del
\t_L \bd \t_L
 -  \del \t_R \bd \t_R - 2 C \ \bd \t_L \del \t_R )$$
$$  - A(\bd \t_R + C \ \bd \t_L)  + \A (\del \t_L +   C  \  \del \t_R) + C A\A
   + \ha b  A{\bd\over \del} A  + \ha \bab  \A{\del\over \bd} \A \  ]  \ .
\eq{7.16} $$
Eliminating $A,\A$ from (7.16) and dropping out the non-local terms we get
the following \sm  action
$$ S(r, \t_L,\t_R)= {\k \over 4 \pi } \int d^2 z [ \del r \bd r   +  {(1+
\bab)(C^2 + b)V\inv}  \del \t_L \bd \t_L
 +  {(1+ b)(C^2 + \bab)V\inv }   \del \t_R \bd \t_R   $$
$$  + \    {(1+ \bab)(1+b)C V\inv }   ( \del \t_L \bd \t_R  +  \del \t_R \bd
\t_L )
+    {(1-b\bab)C V\inv  }    ( \del \t_L \bd \t_R  -  \del \t_R \bd \t_L ) ] \
, \eq{7.17}
 $$
where the function $V$ and the dilaton are given by
$$ V\equiv C^2 - b\bab   \ , \ \ \ \  \p=\p_0 - {1\ov 4} \ln V  \ .
\eq{7.18}   $$
In terms of the coordinates $r, \t, \tt$
$$ S(r, \t,\tt)= {\k \over 4 \pi } \int d^2 z [ \del r \bd r   +  G_{\t \t}
\del \t \bd \t
 +  G_{\tt \tt}   \del \tt \bd \tt  $$
 $$   +   G_{\t \tt}    ( \del \t \bd \tt  +  \del \tt \bd \t )
+   B_{\t \tt}    ( \del \t \bd \tt  -  \del \tt \bd \t ) ] \ , \eq{7.19}
 $$
$$  G_{\t \t}  =
  (C-1) [(1+ \bab)(C- b)  +   (1+b)(C-\bab)]V\inv       \ , \eq{7.20} $$  $$
     G_{\tt \tt}  =
 (C+1) [(1+ \bab)(C+ b)  +   (1+b)(C+\bab)]V\inv        \ ,  \eq{7.21}  $$
$$  G_{\t \tt}  = {(\bab-b)(C^2 -1) V\inv }
 \ , \ \ \  \  B_{\t \tt}   =2 {(1-b\bab)C V\inv} \   \ . \eq{7.22}   $$
If we identify $\a'$ with $  1/\k$ as in (7.11)
  (so that $G_{MN}$ and $B_{MN}$  are to be multiplied  by 1/4)
then in  the bosonic case $(b=\bab = 2/(k-2)=2\a')$ we get
$$   4ds^2 =  dr^2 +   2(1+b)\big[  {C-1 \ov C+b } d\t^2
+  {C+1 \ov C-b } d\tt^2 \big] \ , \ \ \  B_{\t \tt}  ={(1-b^2)C \ov 2(C^2 -
b^2) }  \ , \eq{7.23} $$
while in the heterotic one  ($b=2/k=2\a', \ \bab=0$)
$$ 4 ds^2 =   dr^2 + (1+ bC^{-2}) d\t_L^2
 + (1+b)  d\t_R^2   +  2(1+b)C^{-1}    d\t_Ld\t_R $$
$$  =dr^2 +  2{C-1\over  C}[1 + b{C- 1\over 2 C}]   d\t^2
+  2{C+1\over  C}[1 + b{C+ 1\over 2 C}]   d\tt^2    - 2 b{C^2-1\ov C^2} d\t
d\tt
 \ , \eq{7.24}  $$
$$    B_{\t \tt}  = {1\ov 2C}    \ . \eq{7.25} $$ The  backgrounds (7.23) and
(7.24),(7.25)
coincide in the semiclassical ($b=0$) limit and  are related to (7.11) and
(7.13)  by the coordinate
transformations  $$ z= C= \cosh r\  , \ \ \ \  t=  i [ \ha (1+b)]^{1/2}\t \ , \
\ \ \ x= i [ \ha
(1+b)]^{1/2}\tt \ ,
 \eq{7.26} $$
and
$$ z= C= \cosh r \  , \ \ \ \  t=   {i\ov \sqrt 2}  \t \ , \ \ \ \ x=  {i\ov
\sqrt 2}  \tt \ .
 \eq{7.27} $$
The  metric (7.24)  has rather peculiar `heterotic' (left-right asymmetric)
form.
For comparison, let us  recall  the exact form \sft\bs\ of the `charged black
string'
background \horne\ (corresponding to $[SL(2,\IR)\times \IR ]/\IR $
gauged WZNW theory) represented in the same coordinates $r,\t,\tt$
$$   4ds^2 =  dr^2 +   2(1 + p + b)  {C-1 \ov C+ p + b } d\t^2
+  2(1-p + b) {C+1 \ov C + p -b } d\tt^2  \ ,  \  \eq{7.28} $$
$$
  B_{\t \tt}  ={(1-p_0^2)C \ov 2(C + p_0) }  \ , \  \
\ \ \p=\p_0 - {1\ov 4} \ln [(C+p-b)(C+p+b)]  \ ,    \ \ \ p = p_0 +b \ , \ \
p_0=1 + \s^2
\ ,
 \eq{7.29} $$
where $\s$ is the parameter  that  governs the embedding of the subgroup (for
$\s=0$ (7.28),(7.29)
reduces to the exact $D=2$ black hole background \dvv).\foot {To derive
(7.28),(7.29) one is to add to the $SL(2,\IR)/\IR $ GWZNW action
an extra scalar scalar term $(\del y - \s A)(\bd y + \s \A)$ and then fix $y=0$
as a gauge. The resulting effective action will be (7.16) with $CA\A$-term
replaced by
$(C+p)A\A$ (cf. (4.5)). } Note also that the $SL(2,\IR)$ group space background
is (see (7.15))
$$   4ds^2 =  dr^2 +   2  {(C-1 ) } d\t^2
-  2(C+1) d\tt^2  \ ,  \ \ \  B_{\t \tt}  = \ha C  \ , \ \p=\p_0 \ .
\eq{7.30}
$$

\newsec{Concluding remarks}
There exist five  distinct   classes of  conformal
$\s$-models  associated with gauged or chiral gauged WZNW models:
(1)  models corresponding  to  bosonic   $G/H$ gauged WZNW theories;
(2)  models corresponding to bosonic  $G/H$ chiral gauged WZNW theories;
(3)  models corresponding to (1,0) supersymmetric $G/H$ chiral gauged WZNW
theories;
(4)  models corresponding to (1,1) supersymmetric $G/H$ gauged  WZNW theories;
(5)  models corresponding to (1,1) supersymmetric $G/H$  chiral gauged WZNW
theories.
For  all these  models  the exact dependence of the couplings (background
fields)
on $\a'$  is known and is different in different classes.
The  background fields in the first two  classes depend non-trivially
on $\a'$ (contain terms of all orders in expansion in $\a'$).
There is only one $O(\a')$ term in the  metric in the third class
while the dilaton and the antisymmetric tensor are $\a'$-independent (i.e.
semiclassical).
The fields in the last two classes do not depend on $\a'$.
The backgrounds of the  first and fourth  classes  coincide  in the $\a' \to 0$
limit
(the same is true  for the backgrounds of  the  second, third and fifth
classes).  The $\s$-models  associated with  gauged (chiral gauged) WZNW
theories have the
configuration space of dimension equal to $\dim G/H$ ($\dim G$).
In the case of the abelian subgroup $H$ the models of the second (fifth) class
are equivalent
to a particular  subset of models (axially gauged $(G\times H/H$ with a special
embedding of $H$)
in the  first (fourth) class. The heterotic string solutions are represented by
the models of the
third class (and also  by the  models of the fourth class `embedded' into the
heterotic string theory
by introducing an extra gauge field background \GRT).


The conformal invariance of the simplest  ($D=2$) $\  SL(2,\IR)/U(1)$  model of
the first class was
checked explicitly  to $\a'^3$   \ts (and $\a'^4$ \Jack ) order.
 In \sfts\ we have  checked that the  simplest ($D=3$)  background (7.11)  of
the second class
  solves the  \sm conformal invariance conditions in the $\a'$ and $\a'^2$
approximation.
Though there are  no  doubts that the heterotic background (7.12),(7.13)  of
the third class
solves the corresponding
heterotic \sm  conformal  invariance conditions (or, equivalently, the
heterotic string effective
equations) it may be of interest to check this directly in the $\a'^2$
approximation.
 There exists a scheme  \met\ in which  the
$G,B,\p$--dependent part of the $\a'^2$ term  in the heterotic string effective
 action is  given by
one half of the $\a'^2$ term in the  bosonic  string effective action plus the
non-covariant
contribution $\a'H(\o R - {2\ov 3}  \o\o\o)$ coming from the Lorentz
Chern-Simons modification of
$H=dB $. In contrast to the bosonic case,  there   is no
(modulo  a field redefinition) explicit $O(\a'^3)$ term in the heterotic string
effective
action. Though  the metric (7.13) contains only the two-loop correction, this
does not of
course imply  that the string equations will be  automatically satisfied to all
higher orders.


The backgrounds  corresponding to the chiral gauged $G/H$ WZNW theories
 discussed  in this paper may
be of interest from the point of view of a possible cosmological or black hole
-- type
interpretation.  There are two cases when the resulting space-time metric has
the physical
signature.  If the group $G$ is compact then according to the equivalence
relation \euqiv
we   can  get   just one
time-like coordinate  if  a
compact subgroup $H$ is one-dimensional, i.e. is  $U(1)$.  If $G$ is
non-compact,
but $H$ is compact,  one can  consider the non-compact coset
$G_{-k}/H_{-k}$ and require that the corresponding $\s$-model has just one
time-like coordinate. Since
$H$ appears in  \euqiv  with level ${k-2{\rm g_H}} $  we will get  no
additional
time-like coordinates as long as the condition  $k > 2{\rm g_H}$ is
satisfied.\foot {This counting
argument apparently does not depend on
 whether  the `auxiliary' fields are integrated out or not. All single
time coordinate models characterised by $G/H$ cosets based on simple as well as
direct product
non-compact groups are classified in \IBCS\gin. The complete list of chiral
gauged WZNW theories
with one time-like coordinate is:
 $SU(p,q)/SU(p) \times SU(q) , \ \ SO(p,2)/SO(p),\ \
Sp(2p,\IR)/SU(p) ,\ \ SO^*(2p)/SU(p),\ \ E_6/SO(10)\ ,\ \ E_7/E_6$.
One can also take direct products of the above models with any WZNW
or GWZNW theory with no time-like coordinates. The lowest-dimensional examples
have   $D=6$, i.e. $SO(2,2)/SO(2)$ CGWZN and $SO^*(4)/SU(2)$ CWZNW (in both
cases it is assumed that the overall coefficient in the action is negative,
i.e. $-k$).}


\bigskip
\bigskip
{\bf Acknowledgments}
\noindent
A.A. Tseytlin  is grateful to A. Giveon, E. Rabinovici  and M. Shifman for
useful discussions  and
would like to   acknowledge also a  support of SERC.

\vfill\eject
\listrefs

\end